\documentclass[oneside, a4paper, onecolumn, 11pt]{article}
\usepackage{times}
\usepackage[left=2cm,top=2cm,bottom=2cm,right=2cm]{geometry}
\usepackage[onehalfspacing]{setspace}

\usepackage{graphicx}% Include figure files
\usepackage{dcolumn}% Align table columns on decimal point
\usepackage{bm}% bold math
\usepackage{epsfig}
\usepackage{subfigure}
\usepackage{graphics}
\usepackage{epstopdf}
\usepackage[english]{babel}
\usepackage{amsfonts}
\usepackage{mathrsfs}
\usepackage{amsthm}
\usepackage{amssymb}
\usepackage{enumerate}
\usepackage{amsmath}
\usepackage{subfigure}
\usepackage{algorithm}
\usepackage{algorithmicx}
\usepackage{algpseudocode}
\usepackage[colorlinks=true, allcolors=blue]{hyperref}
\usepackage{multirow}
\usepackage{booktabs}
\usepackage{lscape}
\usepackage{blindtext}
\usepackage[margin=1cm]{caption}
\usepackage{hypcap}
\usepackage{authblk}
\usepackage{tabularx}
\usepackage{chngcntr}
\usepackage[dvipsnames]{xcolor}
\usepackage{xurl}

\floatname{algorithm}{Algorithm}
\usepackage{natbib}
\setcitestyle{open={(},close={)},semicolon,aysep={,}}
\usepackage{longtable}

\newcolumntype{C}{>{\centering\arraybackslash}X}
\makeatletter

\makeatletter

\providecommand{\keywords}[1]
{
\small	
\textbf{Keywords:} human capital; skills; software; economic complexity; networks
}

\title{Using digital traces to analyze software work: skills, careers and programming languages}
\author[1]{Xiangnan Feng}
\author[2,3,1]{Johannes Wachs}
\author[1,4]{Simone Daniotti}
\author[1,5]{Frank Neffke\thanks{Correspondence can be sent to neffke@csh.ac.at.}}

\affil[1]{Complexity Science Hub, Vienna, Austria}
\affil[2]{Corvinus University, Budapest, Hungary}
\affil[3]{ELTE Centre for Economic and Regional Studies, Budapest, Hungary}
\affil[4]{Utrecht University, Utrecht, the Netherlands}
\affil[5]{Transforming Economies Lab, IT:U Interdisciplinary Transformation University; Linz, Austria}

\begin{document}
\maketitle
\begin{abstract}

Recent waves of technological transformation are reshaping work in uncertain and hard-to-predict ways. However, jobs at the forefront of the digitizing economy offer an early glimpse of these changes and leave rich activity traces. We exploit such traces in tens of millions of Question and Answer posts on Stack Overflow for the creation of a fine-grained taxonomy of software skills to analyze human capital in the global software industry. Constructing a software skill space that maps relations among these skills reveals that real-world software jobs demand highly coherent skill sets and that programmers learn through a process of related diversification. The latter process often  leads to the acquisition of lower-value skills. However, when programmers use Python they preferentially target higher-value skills, offering a potential explanation for Python's successful rise as a dominant general purpose language.

\end{abstract}
\keywords

%\section*{Teaser sentence}

%Digital traces of programmers reveal a task structure of software work that explains coding careers and language popularity 

%Software's increasing role in value creation and innovation across economic sectors has unleashed a global competition over talent. We construct a detailed taxonomy of programming tasks --- based on problems and their solutions posted by an online community of software developers --- to analyze the human capital required in complex coding tasks. This reveals how programmers learn and why some programming languages grow in popularity. 
\section{Introduction}

Knowledge of detailed skills and expertise used at work is central to understanding wages, productivity, and overall labor market dynamics \citep{acemoglu2011skills}. Recent work has transformed our ability to do so by using complex network analysis. This has allowed moving away from descriptions that rely on years of schooling, credentials, and work experience to studies of the relations among concrete skills and the tasks they enable workers to perform \citep{anderson2017skill,alabdulkareem2018unpacking,hosseinioun2025skill}. However, most of these studies focus on the content of jobs, not the human capital of  who performs them. This limits our understanding of how workers learn new skills, which is crucial for helping individuals, firms and educators navigate transforming labor markets in times of rapid technological change. To address this, we focus on a specific type of work that has become increasingly important, is changing fast and can be studied in great detail: software development. Constructing a fine-grained taxonomy of programming skills, we create a software development \emph{skill space} that maps relations among these skills. We use this space to analyze which skills are combined in job ads,  how programmers acquire new skills and what role programming languages play therein. 

Software jobs are particularly interesting, because they represent a fast-changing and critical domain of the digital economy \citep{brynjolfsson2014second,wachs2022geography,juhasz2026software}. Software companies dominate the global ranks of the largest firms by market capitalization.\footnote{Of the seven largest companies worldwide, three (Microsoft, Alphabet, and Meta) focus almost exclusively on software, while two others (Apple and Amazon) derive substantial revenue from it.} On the labor market, demand for programming skills has risen sharply over the past two decades: one in 11 U.S. occupations now requires coding skills \citep{onet}, and software-related roles such as \emph{data scientist}, \emph{information security analyst}, and \emph{computer research scientist} rank among the fastest-growing occupations in the US \citep{bls}. At the macro-economic level, this has sparked intense global competition for talent, prompting  major economies to designate software development as a strategic priority \citep{draghi2024future,kazim2021innovation,singh2024study}. 

However, we currently lack a detailed understanding of the changing human capital that is employed in this sector. As new hardware, programming languages, and specialized libraries reshape computing, fresh roles (e.g., web development, DevOps, data science, artificial intelligence) keep emerging \citep{rock2019engineering,papoutsoglou2019extracting,hemon2020agile} and existing roles transform. Existing databases linking occupations to skills \citep[e.g.,  ][]{onet}) are often curated by human experts, making them relatively coarse and slow to update.\footnote{For example, the only software-related skill in O*NET is undifferentiated ``programming'' and the listed software-related work activities often focus on broad domains of application (e.g., ``geoinformatics'' or ``biostatistics''), team dynamics (e.g., managing or collaborating with software engineers), or acquiring software products.} More dynamic data sources are challenging in their own right. For instance, job ads --- frequently used to study job contents --- describe job requirements of ideal-type workers, not human capital endowments of actual workers. Likewise, although rich data from platforms that mediate online freelancing work provide insights into how complex skill interdependencies affect wage premiums \citep{anderson2017skill,stephany2024price}, the extent to which these findings generalize to traditional labor markets remains uncertain.

We address these gaps by analyzing millions of posts from Stack Overflow (SO)---the largest question-and-answer forum for programmers \citep{anderson2012discovering}. Each post describes a specific technical challenge and includes tags that indicate the tools, languages, or concepts involved \citep{barua2014developers}. These challenges reflect real-world coding tasks and  by posting answers users demonstrate skills and expertise in these tasks. Indeed, SO postings have real-world value as signals  to prospective employers of a programmer's capabilities \citep{xu2020makes}. 

Interpreting tags as a structured, albeit partial, description of posted questions, we aggregate questions to create a software-skill taxonomy by identifying clusters of frequently co-occurring tags. We then introduce a notion of \emph{relatedness} among programming skills based on co-occurrence patterns in users' posting histories. We use this relatedness to construct a skill space where skills are connected if they are often jointly possessed by the same users. We then use this taxonomy and skill space to  examine the postings and learning trajectories of programmers. 

We find that our skill constructs can be used to predict salaries in real-world job postings. Notably, wages vary greatly across skill profiles, with higher-paying jobs concentrated in more specialized or in-demand fields such as AI and machine learning. Similarly, relatedness estimates derived from interactions with the SO platform predict which combinations of skills are demanded in real-world job ads. This not only shows that our taxonomy captures meaningful structures in the software development landscape, but also that software development jobs are coherent in the sense that they often require sets of closely related skills.

The skill taxonomy also sheds light on the role and evolution of programming languages. Programming languages differ in the expertise they support. Analyzing which skills are used with which programming languages, we observe a ``nested'' pattern reminiscent of similar patterns observed in ecology \citep{bascompte2003nested}, economic development \citep{hidalgo2009building,tacchella2012new,mariani2019nestedness} and skill hierarchies \citep{hosseinioun2025skill}. This means that although programming languages may have been tailored for specific use cases, this is not observed in how programmers use languages in practice. Instead, we observe that skills have an implicit hierarchy such that the most ubiquitous skills are supported by almost all languages, whereas rare skills often rely on the most versatile ones. A prime example of the latter is the  programming language \emph{Python}, a popular language that has seen significant growth in recent years \citep{economist2018python}. Our skill taxonomy shows that this growth has been accompanied by changes in the way Python is used, with Python becoming increasingly versatile, rapidly increasing the variety of skills it supports to become today's premier general purpose language. 

At the individual user level, we observe two patterns. First, developers excel ---as inferred from community feedback on their posts--- at the skills they are most active in. Second, developers acquire new skills over their SO careers, and the relatedness network strongly influences which skills they learn: developers are fifteen times more likely to acquire skills related to skills they already master than to unrelated ones. At the same time, we find that developers usually add skills with lower market value, which are presumably easier to learn. However, Python enables qualitatively different learning paths. Specifically, unlike users of most other languages, Python users often branch into more, not less valuable skills. This observation could help explain Python’s success, despite its not being the oldest or fastest language: its ecosystem may make it easier for developers to develop higher-value expertise.

Our analysis contributes to the growing literature that analyzes the content of work. This literature has shown that as wages and careers have come to depend on how workers specialize \citep{hosseinioun2025skill}, interact \citep{deming2017growing}, complement one another \citep{neffke2019value}, and move across places and jobs \citep{frank2024network}, fine-grained skill taxonomies are increasingly valuable. Such taxonomies clarify how workers transition between occupations~\citep{del2021occupational}, how skills substitute for or complement each other~\citep{anderson2012discovering, neffke2019value}, and how occupations evolve over time~\citep{nedelkoska2021eight}. 
This line of research uses data on skill requirements and work activities extracted from large-scale surveys \citep{onet}, expert assessments \citep{european2017esco}, job ads \citep{borner2018skill} or online freelance platforms\citep{anderson2017skill,stephany2024price} to describe jobs as networks of skills or tasks. We contribute to this debate by focusing away from the task requirements listed in descriptions of jobs toward the human capital and skills of the people who fulfill them. This allows us to study how workers specialize and learn new things, processes critical to a successful career.

%Focusing on the software sector, we propose a new approach that uses data from Stack Overflow (SO), a very large, online question-answer database. The core insight we leverage is that questions on SO describe coding problems that programmers encounter in their work. As such, they can be understood as instantiations of software-development job-tasks and users who provide answers to these questions implicitly demonstrate skills or expertise required to tackle these tasks. This suggests that we can use the questions in SO to construct a taxonomy of canonical software skills that is highly granular, flexible and based in data. This taxonomy allows us to describe the human capital endowments of people, as opposed to the human capital requirements of jobs. That is, we analyze what software developers know, not what they are supposed to know according to their job descriptions. This allows us to analyze how programmers learn new skills and what the value of these skills is. Moreover, programmers can choose from a wide menu of programming languages and we show that programming languages differ in the type of software developers they support. However, the mapping of software development skills to programming languages changes over time, as illustrated by the rapid rise of Python as a general purpose language. Our skill taxonomy shows that this rise exploits a remarkable strength of Python: it allows programmers to more quickly transition to valuable specializations in programming.   
\section{Mapping software work}

In this section, we describe how we construct a software skill taxonomy and a space of software skills. First we describe the raw inputs: posts on Stack Overflow and their tags, labels which capture post content. We then explain how we use a network-based clustering algorithm, the stochastic block model (SBM, see Appendix~\ref{sec:app_SBM}), to identify clusters of tags which frequently co-occur in posts. These clusters are our skills, from which we can derive descriptions of user skills and skill values. Finally we demonstrate how skills can be embedded in an abstract space, in which related skills are near one another.

\subsection{Data}\label{sec:data}
Our main dataset is the corpus of questions and answers (together ``posts'') on Stack Overflow from the site's launch in August 2008 to June 2023. The 23 million questions in this dataset are labeled by tags from a curated system of about 66,000 tags referring to software concepts and technologies such as \emph{key-value}, \emph{scipy} and \emph{svm}. For each tag, SO provides a textual description, typically consisting of a few sentences. Tags are heavily moderated by humans and automated processes, and are deduplicated through a dictionary of synonyms. We focus on tags that are used at least 1,000 times. We split tags into a two groups: 5,083 general tags and 282 tags that refer to programming languages, defined broadly. For the latter, we merge different versions of the same programming language, converting, for instance \emph{python-3.x} into the generic tag \emph{python}. When we define skills, we use all available posts, with answers inheriting the tags of the questions they answer. When we analyze users, we only use their answer posts (ignoring the questions they post) to assess the content of their human capital. To ensure sufficiently accurate descriptions of this human capital, we restrict our analysis to users who post at least 10 answers over the entire period our data cover.

Stack Overflow also provides access to yearly surveys of its users, the \emph{Stack Overflow Developer Survey}. This provides anonymous, self-reported information on salaries, as well as on tools and languages used that can often be linked to tags in SO. We use the survey data from year 2023, which contains around 12,000 U.S. participants who report  salary information. 

Finally, we scraped job advertisements from ``hnhiring.com'', an index of jobs from the website\footnote{\url{https://www.hackernews.com/}} Hacker News' \emph{``Who is Hiring?''} posts. This yielded a total of 50,998 software related job advertisements. We use approximately 46,000 ads to study skill requirements of jobs and 3,949 ads listing salary information to test the ability of skill requirements to predict wage offers.

\subsection{Constructing a taxonomy of software skills}\label{sec:taxonomy}
Our goal is to construct a taxonomy of skills related to tasks in software development work. Software development tasks can be viewed as recurring categories of programming problems that developers encounter. The ability to provide advice on how to tackle such problems demonstrates skills and expertise in the area. Therefore, SO, with its vast repository of user-generated questions and answers, offers a unique window onto these real-world challenges and the human capital required to meet them. 

In order to organize its community and facilitate information retrieval, the SO platform labels each question with one or more of 66,000 curated tags that indicate the specific tools, frameworks, or concepts involved. Tags are deduplicated and moderated by automated bots and human moderators, ensuring relatively high quality. Before constructing a skill taxonomy, we filter tags in two ways. First, we drop tags appearing fewer than 1000 times. Second, we distinguish tags that refer to programming languages rather than concepts or methods. We drop these tags from our skill taxonomy, because we programming languages are rather tools to implement software skills; the same skill could be applied in different languages. While languages themselves are specialized \citep{valverde2015cultural,juhasz2026software}, they are a different category. In fact, we will later analyze the relationship between skills and languages used.

To identify tags that refer to programming languages, we rely on wiki pages, which are available for most common tags. We first filter the pages to find tags that include ``language'' or  ``framework'' among their keywords. We then manually review these tags to select those that refer to programming languages, yielding a total of 282 programming languages.

Some programming tags refer to specific versions of a language, such as ``python-3.x'' or ``python-3.7''. In these cases, we merge all tags referring to different versions of the same programming language. Yet, there is some ambiguity as to what counts as a programming language (e.g., web frameworks and developer environments, which are not fully fledged programming languages). We remove such tags when analyzing the relation between skills and languages, yielding 114 unambiguous instances of programming languages.

\subsubsection{Identifying software skills}

With the processed set of tags in hand, we can now identify canonical software skills in our data set. To do so we express the relation between questions and their tags as a bipartite network with 18,154,593 question nodes that connect to 5,083 tags through 33,378,590 edges. We use these tag–question relationships to group related tags into ``canonical'' areas of software development expertise (``skills''), applying an SBM \citep{peixoto2017nonparametric,peixoto2014hierarchical,gerlach2018network} to detect communities in the bipartite graph that connects questions to tags (Fig.~\ref{fig:fig1_skilldef}a). SBMs essentially estimate a coarsened version of a network. To do so, they identify communities of nodes and estimate parameters that yield parsimonious descriptions of edge distributions within and between communities. To sharpen the boundaries of inferred communities, we remove $20\%$ of tags that are only weakly connected to the other tags in their community (see Appendix~\ref{sec:app_SBM}). This yields 247 different communities. Dropping communities with fewer than 3 tags, we retain 237 communities, connected to 4,054 tags, which we will henceforth will refer to as ``skills''.

To generate concise names for each skill community, we label them by an example of a programming skill that illustrates the tags in the community. That is, we name each skill after the task it allows to perform. To produce these labels, we prompt the large language model ChatPGT-4.0 as follows:

\begin{quote}
    I have a list of tags related to software engineering and computer science. The tags are ordered by their importance, listing the most important task first. Using no more than 10 words, could you please suggest a general task related to the following tags 'task\_description' as key words? Please make sure that the task is  general and not too specific. Please only return json content and not anything else.
\end{quote}

This process yields labels that capture the essence of the tags comprising each group (see Appendix~\ref{sec:skill_space_labels}), such as  \emph{develop NLP models and tools for text analysis}, \emph{establish secure server access and control}, and \emph{deploy and secure a scalable web app}.

\subsection{User skill vectors}
Skills are rarely valuable in isolation \citep{neffke2019value}. Therefore, jobs often require multiple skills. When skills complement each other, they are often required in the same jobs \citep{anderson2017skill,alabdulkareem2018unpacking,stephany2024price}. We investigate this skill bundling by identifying which combinations of skills are frequently mastered by the same SO users. To assess to what extent a user masters a skill, we rely only on the answers they post, not on the questions they raise.

We use the skills (defined from the tasks they allow  performing) extracted from the question-tag network to describe the activity of individual SO users as a 237-dimensional vector. Entries of this vector count how many answers the user has provided to questions that are labeled with tags that are associated with each corresponding skill. Specifically, we express the user's expertise in time period $t$ as:
\begin{equation}\label{eq:skillvector}
   \vec{S}_u^t = \begin{bmatrix}
X_{u,1}^t  \\
... \\
X_{u,\theta}^t \\
... \\
X_{u,N_\theta}^t  \\
\end{bmatrix}, 
\end{equation}
where $N_\theta=237$ is the total number of skills in our taxonomy and $X_{u,\theta}^t$  the number of answers posted in period $t$ by user $u$ to questions labeled with a tag belonging to the community of tags associated with skill $\theta$. We will interpret this number as the user's expertise intensity in a skill.

\subsection{Skill values}\label{sec:skill_value}
We estimate the value of a skill based on salary information from the Stack Overflow developers survey. Because wages differ widely across countries, we focus on 12,000 US respondents that report a salary in 2023. Respondents list the technologies they use, which often correspond to tags in the SO platform. We use tags that are used in both the survey and the general SO database to find closely matching survey respondents for each SO user. Selecting the 300 most similar survey respondents, we calculate the weighted average salary reported by these respondents. We will use this average as a rough estimate of the wage that the user could command on the US labor market:   

\begin{align}\label{eq:user_value}
    V_u &= \sum_{\rho \in R_u} \frac{M_{u,\rho}}{\sum_\alpha M_{u,\alpha}} V_\rho,
\end{align}
where $V_u$ is user $u$'s inferred US wage, $V_\rho$ the wage of survey respondent $\rho$, and $R_u$ the set of survey respondents we matched to user $u$. Weights, $M_{u,\rho}$, are given by the number of tags that user $u$ and respondent $\rho$ share. 

To assign a value to a skill, we next calculate the weighted average salary of users possessing the skill, using the  skill's intensity in users' skill vectors for the years 2018-2023 (the period in which we collect job ads) as weights:

\begin{align}\label{eq:skill_value}
    V_\theta &=  \sum_u \frac{X_{u,\theta}^{2018, 2023}}{\sum_{u'} X_{u',\theta}^{2018, 2023}}  V_u.  
\end{align}

\subsection{Constructing a skill space}
To determine the relatedness between skills, we analyze which skills are often possessed by the same users. To do so, we collect the skill vectors of all users in an 
$(N_
\theta \times N_u)$ matrix $\textbf{S}$, with $N_u$ the total number of users in the dataset and $N_\theta = 237$ the total number of skills. Next, we ask how often two skills co-occur within the same users by multiplying $1(\textbf{S}>0)$ with its transpose: $1(\textbf{S}>0)1(\textbf{S}>0)'$, where $1(.)$ is an indicator function that evaluates to 1 if its argument is true. Finally, we calculate the Pointwise Mutual Information ($\mathrm{PMI}$) of probability $p_{\theta,\kappa}$ that skill $\theta$ and $\kappa$ co-occur in the same user to capture the amount of (information-theoretic) surprise involved in observing the joint probability $p_{\theta,\kappa}$ against a benchmark where draws of skills $\theta$ and $\kappa$ are independent:

\begin{equation}
\mathrm{R}_{\theta,\kappa} = \mathrm{PMI}_{\theta,\kappa} = \log \frac{p_{\theta,\kappa}}{p_\theta p_\kappa},
\end{equation}
where $p_\theta$ and $p_\kappa$ are marginal probabilities, i.e.,  $p_\theta = \sum_\gamma p_{\theta,\gamma}$ and  $p_\kappa = \sum_\mu p_{\mu,\kappa}$. We estimate $\mathrm{PMI}$ values and their credible intervals using a Bayesian statistical framework developed by \citep{van2023information}. We will refer to these values as the \emph{relatedness} between skills. 

\begin{figure}[t!]
    \centering
    \includegraphics[width=17cm]{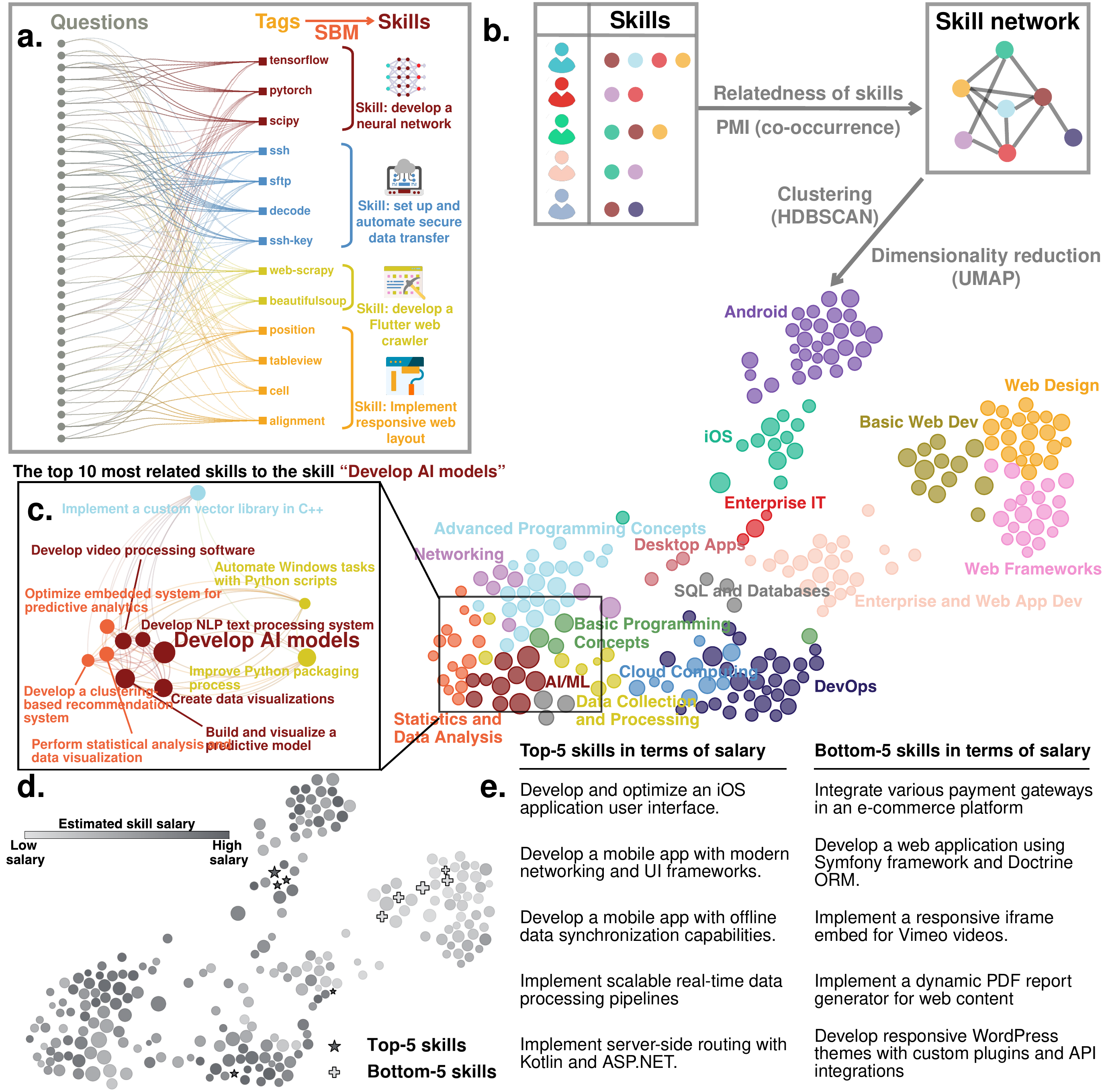}	
    \caption{Mapping software skills. \textbf{a.}  Stylized depiction of the bipartite question-tag network. SBM groups tags into communities (\emph{skills}) that connect to similar sets of questions. ChatGPT-4.0 finds a common label that summarizes each community’s tag information. \textbf{b.} Skill space. Pointwise mutual information (\emph{relatedness}) expresses how surprisingly often two skills are possessed by the same users. UMAP embeds the resulting relatedness network in a 2-dimensional plane (the \emph{skill space}). \textbf{c.} Close-up on the skill to \emph{develop AI models},  depicting the original network structure among the 10 most closely related skills. \textbf{d.} Skill values. Nodes are colored according to their skill value, estimated from salary information in the 2023 SO developers survey. Darker shades indicate more valuable skills. \textbf{e.} Table of the five most and least valuable skills.}\label{fig:fig1_skilldef}
\end{figure}

We use these relatedness estimates to construct a \emph{skill space}, a network where nodes represent skills connected by edges that indicate the relatedness between them. In doing so, we only connect skills with positive relatedness values, ignoring negative relatedness (where skills co-occur less frequently than our random null model predicts) by setting all negative $\mathrm{PMI}$ values to 0. To help interpret this skill space, we embed the relatedness network in a 5-dimensional space using UMAP~\citep{McInnes2018}. Next, we cluster sets of contiguous skills using HDBSCAN~\citep{McInnes2017}. These clusters are represented in colors of Fig.~\ref{fig:fig1_skilldef}b. The node coordinates in this figure are determined by a UMAP embedding of the relatedness network in a 2-dimensional space. Note, however, that neither the clusters nor the visual representation of the skill space will play a role in our quantitative analyses.

Clusters often group skills with comparable labor market returns. This is illustrated in Fig.~\ref{fig:fig1_skilldef}d, where, we plot the inferred skill values of eq.~(\ref{eq:skill_value}). For example, on the right side of the skill space, we find three clusters revolving around website development that are typically linked to lower wages. By contrast, skills in the top center---focusing on advanced iOS and Android development---offer significantly higher earnings. Additional high-value clusters include those for AI and machine learning (\emph{AI/ML}), \emph{Advanced Programming Concepts}, \emph{Cloud Computing}, and \emph{DevOps}. The table of Fig.~\ref{fig:fig1_skilldef}e lists the five highest-paying skills, three of which are related to iOS app development, and the five lowest-paying skills, all associated with more basic web development. 

Notably, clusters that are conceptually linked---such as \emph{AI/ML} and \emph{Statistics and Data Analysis}, and \emph{iOS} and \emph{Android} development---tend to lie near each other in the skill space. To further investigate the network structure of the skill space, we can examine individual skills in greater detail.  For example, the ten skills most closely related to \emph{developing AI models} (Fig.~\ref{fig:fig1_skilldef}c) reveal strong connections to other skills in statistics, as well as skills related to image and natural language processing, data visualization, and packaging models into standalone applications. Fig.~\ref{fig:user_share_change} shows changes in the relative importance of different skills, highlighting important areas of growth around skills in web frameworks, but also in high-value skills related to AI, cloud computing and Android.   

\section{Results}

\subsection{Validation in job ads}
How well do these software skills capture real-world development work beyond SO? To find out, we examine all 50,998 job ads posted to the Hacker News website between January 2018 and May 2024. These ads are unstructured job descriptions, sometimes including salary and expected work activities (tasks). We prompt an LLM (ChatGPT-3.5) to extract salary information and work activities (see Appendix~\ref{sec:job_ad_example}), which yields about 46,000 jobs listing at least 3 work activities, and 3,949 jobs that contain at least 1 work activity as well as salary information. When ads describe more than one position, the LLM provides information on each job separately.

\begin{figure}[htbp]
    \centering
    \includegraphics[width=17cm]{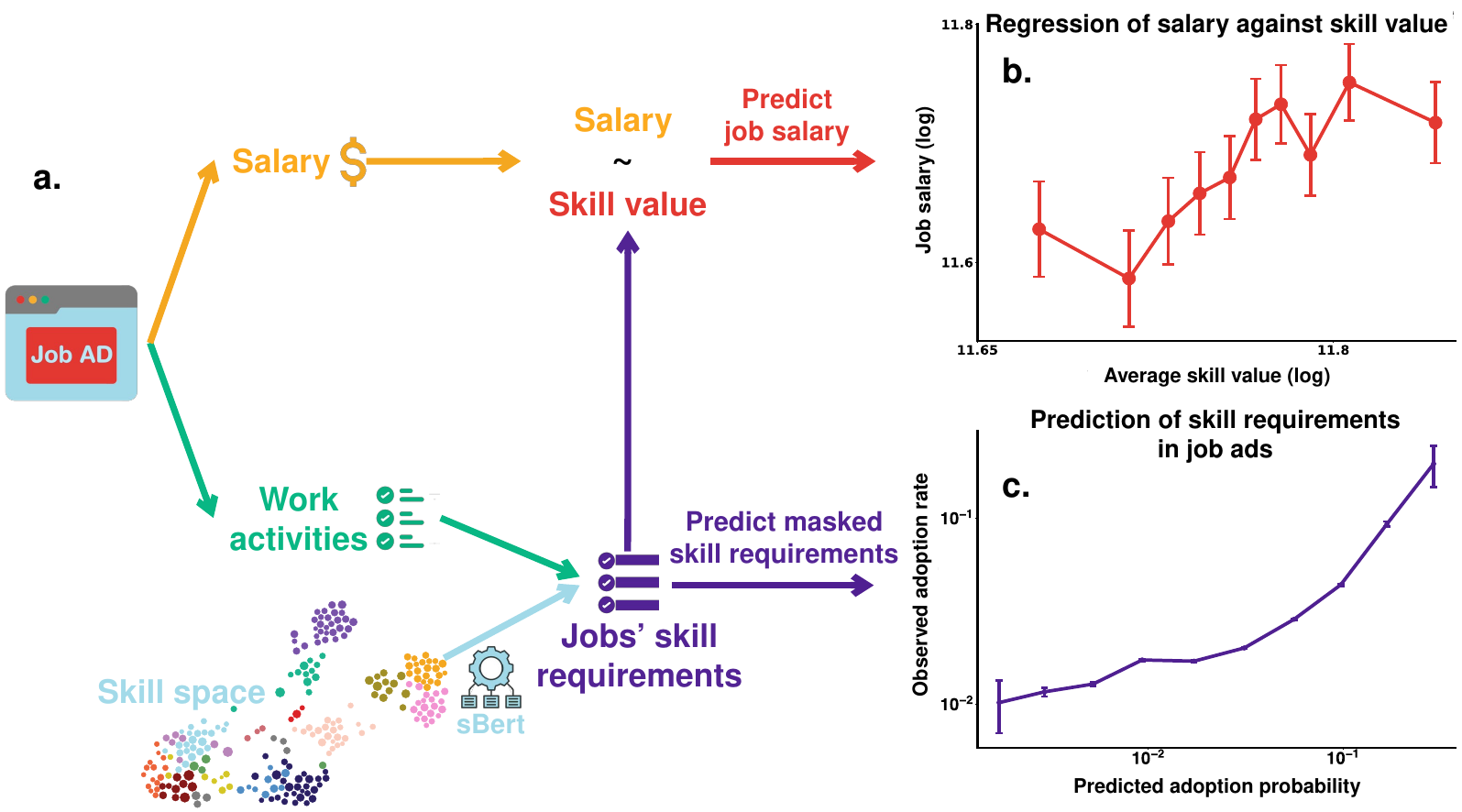}	
    \caption{Job ads. \textbf{a.} Schematic representation of the workflow to extract salary and work activities from online job ads by prompting ChatGPT. Work activities are converted to the 237-dimensional skill requirement vectors based on cosine similarities between text embeddings of work activities and skills. \textbf{b.} Prediction of wage offers. Jobs are grouped into equally sized bins based on the average value of required skills. The vertical axis shows the estimated mean of the advertised wage offers. \textbf{c.} Prediction of skill requirements. 40\% of  elements in each skill requirement vector are masked and grouped into 10 equally sized bins based on the average proximity of a masked skill requirement to all unmasked skill requirements in a job. The plot displays the estimated probability that a specific masked skill in a bin is required for a job.  Vertical bars in panels b and c represent 95\% confidence intervals.}\label{fig:fig2_jobads}
\end{figure}

Next, we match each extracted work activity to the closest skill in our taxonomy. To do so, we embed work activities and skills in a 384-dimensional vector space using sBERT~\citep{reimers-2019-sentence-bert}. For work activities, we embed the phrases extracted by the LLM; for skills, we calculate the embedding of their labels.\footnote{To increase accuracy, we use somewhat longer labels of up to 30 words generated by ChatGPT-4.0.} Finally, we match each work activity in a job ad to the closest skill based on the cosine similarity between the embeddings of the work activity and the skills. When we cannot find any skill with a cosine similarity of at least 0.3, we drop the work activity. This yields a 237-dimensional vector of zeros and ones that describe each job's requirements in terms of the skills of our taxonomy.\footnote{In Appendix~\ref{section_alternative_match}, we show that our findings are robustness when we change this threshold or when use embeddings for skills based directly on their tags instead of their labels.} We combine this vector with information on skill-values and relatedness among skills. Fig.~\ref{fig:fig2_jobads}a illustrates this process. 

We use these data in two ways to test the validity of our skill taxonomy. Our first validation focuses on the set of advertised jobs that also post salary information. It estimates the association between the average value of a job's skills and the posted wage in the job ad. We find that skill requirements help predict wage offers (Fig.~\ref{fig:fig2_jobads}b). Salaries in job ads increase alongside the average value of the skills a job requires. Regression analysis confirms this: a 10\% rise in the average value of these skills is associated with a 9\% increase in the wage offer for the advertised job (see Appendix~\ref{section_salary_regression}).

Second, we ask whether we can predict the skill composition of a job. To do so, we take the skill vectors associated with each job ad and stack these vectors into a dataset that contains for each combination of a skill and an advertised job whether or not the skill is required in the job. Next, we mask 40\% of observations in this dataset and calculate the average relatedness of a masked skill requirement, $\theta$, to all unmasked skill requirements in the same job, $j$:
\begin{equation}\label{eq:density_job}
\mathrm{D}_{\theta, j} = \bm{\bar{\mathrm{r}}}_{\theta}^j \bm{\mathrm{s}}^{j},
\end{equation}
where $\bm{\mathrm{s}}^j$ is a vector that subsets the unmasked (i.e., ``visible'') portions of job $j$'s skill requirement vector, $S_j$.  $\bm{\bar{\mathrm{r}}}_{\theta}^j$ refers to the corresponding row-normalized vector of relatedness matrix $\mathrm{R}$, consisting of elements  $\frac{\mathrm{R}_{\theta,\zeta}}{\sum_{\tau\in \Xi_j} \mathrm{R}_{\theta,\tau}}$, where $\Xi_j$, the set of unmasked skills in the skill requirements vector of job $j$, such that $\zeta \in \Xi_j$ and $\theta \notin \Xi_j$.

The skill space strongly predicts which skill requirements appear together in job ads. To show this,  Fig.~\ref{fig:fig2_jobads}c plots how the estimated probability that a job requires a particular (masked) skill changes with that skill's density in the job's unmasked skill requirements. First, we sort all masked observations by the unmasked skill's density in the job, then bin them in 10 equally sized bins and calculate relative frequencies. Next, we plot these estimated probabilities on the vertical axis against the bin's average skill density on the horizontal axis. 95\% confidence intervals are approximated with a normal distribution, using the following estimate of the standard error of the mean:
\begin{equation}\label{eq:sem_diversification}
    \hat{\sigma}_b = \frac{1}{\sqrt{n_b}}\sqrt{\hat{\pi}_b(1-\hat{\pi}_b)},
\end{equation}
where $n_b$ the number of observations in bin $b$ and $\hat{\pi}=\frac{d_b}{n_b}$ (where $d_b$ the number of skills required in jobs in bin $b$) representing the relative frequency with which masked skills are required in bin $b$.

The probability that a masked skill is required (i.e., has a ``1'' in the skill requirements vector) increases sharply with its relatedness to the unmasked skill requirements, rising almost 20-fold from 1.0\% for unrelated skills to 19.6\% for strongly related ones. This not only shows that our SO-based skill taxonomy is meaningful  in the real-world software development labor market, but also that software development jobs are coherent: they, on average, require skill combinations that are often mastered by the same users on SO.   

\subsection{Mapping skills to programming languages}

Programmers can draw on a wide variety of programming languages, each with different levels of suitability and popularity for specific use cases. However, which languages best serve which areas of software development remains hotly debated. Our skill constructs promise a novel way to approach this question: using tags that refer to specific programming languages we can associate SO posts with the languages they refer to (see section~\ref{sec:taxonomy}). This, in turn, allows us to track which skills are used with which languages.

\begin{figure}[htbp]
    \centering
    \includegraphics[width=17cm]{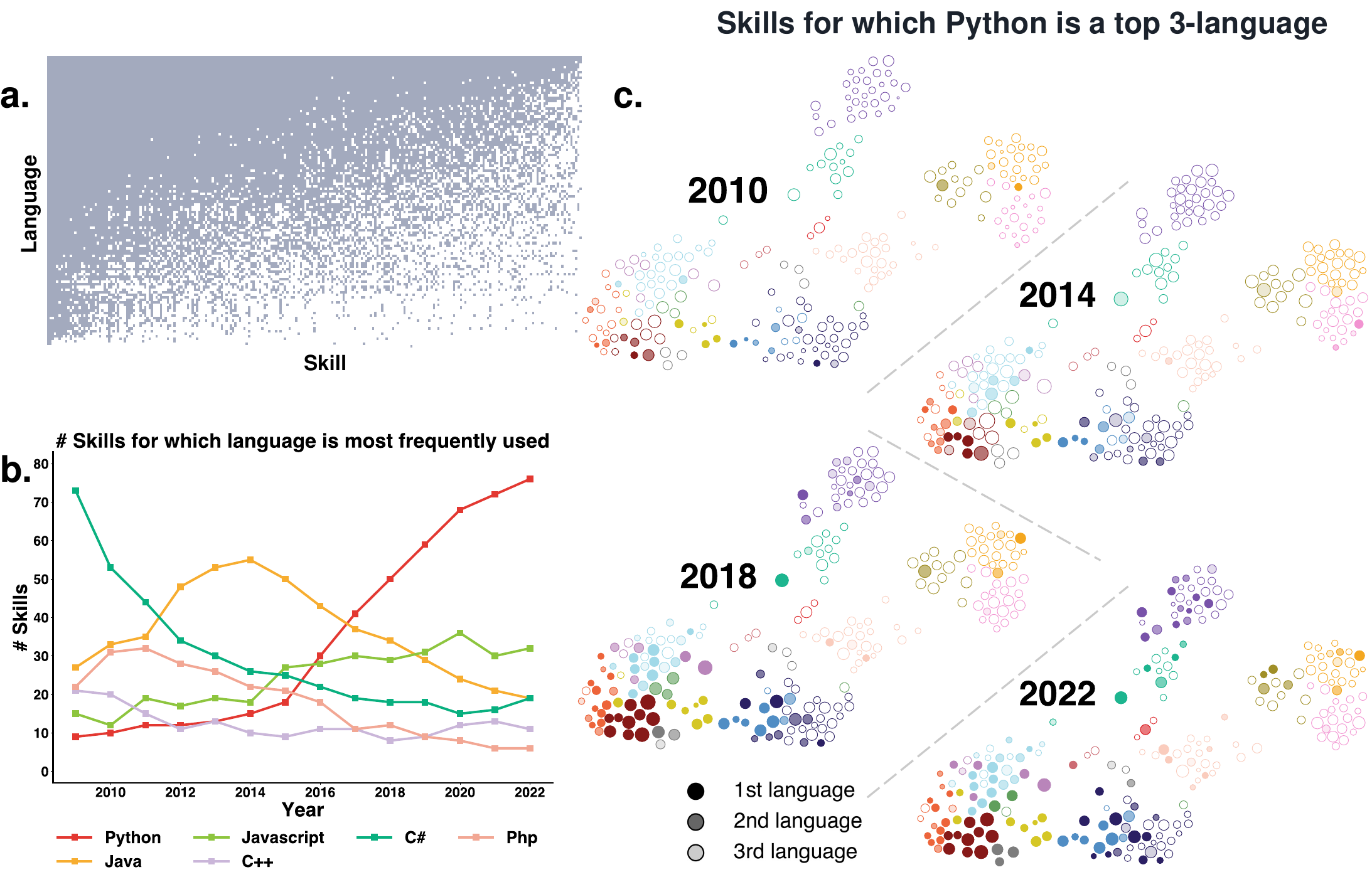}	
    \caption{Programming languages. \textbf{a.} Skill-language matrix. Elements are colored when at least 10 SO users have at least one answer post in the skill-language combination. \textbf{b.} Number of skills for which a programming language ranks as the top language in terms of SO users. The graph shows time-series for the six largest languages in terms of cumulative SO posts between August 2008 and June 2023. \textbf{c.} Python's skill footprint. Nodes are colored if Python ranks among the top 3 programming languages for the associated skill. Fully colored:  1st rank, 50\% transparency: 2nd rank, 75\% transparency:  3rd rank.}\label{fig:fig_python}
\end{figure}

Fig.~\ref{fig:fig_python}a presents a matrix illustrating which programming languages are used by at least 10 users with each of the 237 skills we identified, based on answer postings whose tags refer to both the skill and the language. A visual inspection suggests that the matrix is triangular. This indicates that skills are \emph{nested} \citep{hosseinioun2025skill}, as confirmed by an NODF value of 81.5 \citep{payrato2020measuring}. Accordingly,  skills can be roughly ranked from general to specialized based on how widely they are implemented across languages. That is, although different languages may be developed with specific use cases in mind, in practice, this has not led to a strict specialization where dedicated programming languages serve specific areas of software development expertise. Instead,  whereas common skills can be performed in almost any language, highly specialized skills often require  generalist languages.  

Fig.~\ref{fig:fig_python}b illustrates how different programming languages compete for user types over time. Focusing on the six largest languages on Stack Overflow, it tracks how many distinct skill areas each language dominates by user counts. The most prominent trend is Python’s steady climb: from just nine skills in 2009, it rises to become the top language for over 80 of the 237 skills by 2022, surpassing well-established options like C\# and JavaScript. In the Appendix~\ref{section_swift_objc}, we further demonstrate how languages compete for dominance in expertise areas by focusing on the abrupt shift in iOS programming from Objective-C to Swift.

Fig.~\ref{fig:fig_python}c visualizes Python’s growth, highlighting all skills for which it ranks among the top three languages. Initially focused on data science, Python spread between 2014 and 2018 into software development areas that range from app development to web-related work, DevOps, and cloud computing. Today, it is the dominant language for at least one skill in every cluster. As detailed in the Appendix~\ref{section_reweighted}, this pattern is preserved when we re-weight languages by their GitHub script counts --- a proxy for real-world language usage --- indicating that Python’s reach extends beyond Stack Overflow.

\section{User expertise}
We can study the expertise of SO users, by analyzing the 237-dimensional skill vectors of eq.~(\ref{eq:skillvector}), which count how many answers a user provides to questions associated with each skill in our taxonomy. These vectors therewith capture the intensity of engagement in different areas of software work.

\subsection{Voting}
To study whether engagement is related to users' proficiency in a skill, we analyze how many votes their answers receive. Votes are social feedback from other users indicating appreciation for a post. However, before deciding to post an answer, users will first look at existing answers to see if they can improve or add to them. As a consequence, the sample of answers that users post is highly selected. To avoid that this introduces sample selection bias, we limit ourselves to the first recorded answers for each question. Furthermore, users will post answers to question for which they perceive themselves to be sufficiently knowledgeable. We address this type of sample selection bias in Appendix~\ref{sec:SI_causal}, where we exploit exogenous variation in activity patterns throughout the 24-hour daily cycle.

To assess the effect on the upvotes a user's answers receive, we estimate the following regression model:
\begin{equation}\label{eq:votingregression}
y_{a,\theta} = \beta_x \log(X_{\theta,u(a)} + 1) + \beta_a \log(A_{q(a)}) + \beta_v \log(1+\sum_{i \in Q_{q(a)}} v_i) + \eta_a,
\end{equation}
where $X_{\theta,u(a)}$ is user $u(a)$'s expertise in terms of the number of answers provided to questions involving skill $\theta$ in the two calendar years preceding the calendar year in which answer $a$ was provided. The next two variables serve as control variables for the general amount of interest a question generates on SO: the first, $A_{q(a)}$, is the number of answers provided to question $q(a)$. The second term counts the number of upvotes across all answers to the question, where  $Q_a$ is the set of answers provided to question $q(a)$ and $v_i$ the number of upvotes that answer $i$ received.  $\eta_a$ is a disturbance term. Finally, the dependent variable $y_{a,\theta}$ is either a binary variable that indicates whether answer $a$ is a question's top answer or a continuous variable capturing an answer's popularity: $\log(v_a +1)$, the logarithm of the number of upvotes answer $a$ receives, augmented by 1 to avoid $\log(0)$ issues. In some cases,  an answer is associated with multiple skills. Whenever this happens, we duplicate the observation accordingly, such that this answer generates one observation for each skill involved.

We expect that users are particularly apt at answering questions that strongly connect to their prior expertise. %Testing this hypothesis is complicated, since users often decide to answer a question only after gauging solutions posted by others. This introduces selection biases. In Appendix~\ref{sec:alternative_voting}, we address these biases using instrumental variables. Here, we mitigate them partially by focusing on only the earliest answer posted to each question. We then ask whether users with greater prior expertise in skills connected to the question receive more votes. Concretely, we regress a binary outcome---whether the answer is the top-voted---on the user’s prior expertise in a skill (``skill expertise''), controlling for the question’s total number of answers and sum of votes. As a robustness check, we also predict the total votes an answer receives. 
Fig.~\ref{fig:fig_userlevel}a supports this conjecture, showing that there is  a statistically significant, positive link between users’ skill expertise and the popularity of their answers. The estimates plotted in the figure indicate that a 10\% rise in skill expertise boosts the likelihood of an answer becoming the top-voted solution by 0.1 pp ($\pm0.004$) and increases the number of votes it receives by 0.14\% ($\pm0.006\%$). In the Appendix~\ref{sec:SI_causal}, we show that correcting for sample-selection issues increases estimated effects to 0.6 pp ($\pm0.35$) and 0.9\% ($\pm0.41\%$), respectively. Together, these results corroborate the value of skill vectors as summaries of a user's expertise.

\subsection{Skill acquisition}
To study how user learn new skills, we randomly divide the population of users into two equally sized samples. We use the first sample, $U_1$, to construct a skill relatedness matrix from co-occurrence patterns of skills in users. The second sample, $U_2$, is used to estimate how relatedness affects user activity on SO. To do so, we collect information from all answer posts by these users within rolling two-year intervals. That is, we construct expertise vectors $\vec{S}_u^t$, where $t$ denotes an interval that stretches across the two calendar years preceding year $t$. For instance, $\vec{S}_u^{2018}$ refers to the answers user $u$ provides between January 1, 2016 and December 31, 2017. We collect these vectors year-by-year across all users in sample $U_2$ in matrices $T^t$.

Analogously to eq.~(\ref{eq:density_job}), we calculate for all users, $u \in U_2$, and years, $t$, their expertise density around each skill, collecting results in user-skill matrix $D^t$: 
\begin{equation}\label{eq:density_user}
\mathrm{D}^{t} = \bar{ \mathrm{R}}^{U_1} \mathrm{T}^{t},
\end{equation}
where $\bar{\mathrm{R}}^{U_1}$ is the row-normalized relatedness matrix, constructed from the skill vectors of users in sample $U_1$.

Next, we analyze how SO users develop new expertise over time. To do so, we first test whether the relatedness among skills---reflected in the skill space---predicts which new skills a user will acquire. We take all skills in which user $u$ showed no activity in the two calendar years preceding $t$. These skills represent the set of skills that user $u$ has not yet acquired. We tag all such skills with a $1$ whenever the user posts at least one answer related the the skill in year $t$, and with $0$ otherwise. Next, we divide all user-skill combinations where a skill is at risk of being acquired by the user into 10 equally sized bins based on their density values. We then estimate the probability that users acquire a skill in year $t$ as the relative frequency of 1s in each bin. Standard errors are calculated analogously to eq.~(\ref{eq:sem_diversification}).

\begin{figure}[htbp]
    \centering
    \includegraphics[width=17cm]{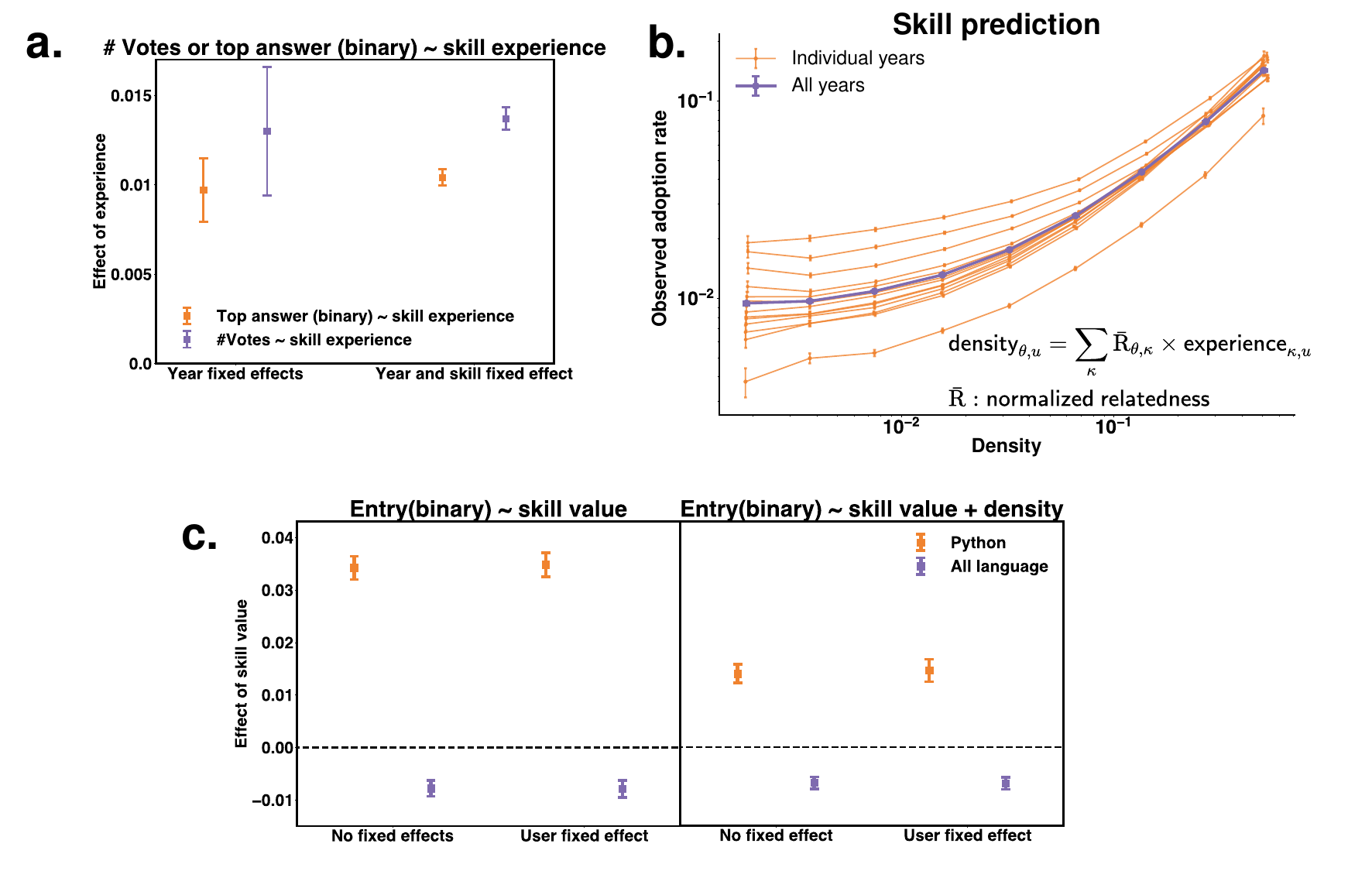}	
    \caption{\textbf{a.} Regression analysis of answer popularity on skill expertise. Popularity is measured as the number of votes the answer receives, $\log (\# \text{votes}_{a} + 1)$ or whether or not the answer is the top answer to the question, skill expertise as the number of prior answers the user provided to questions requiring skill $\theta$ in the preceding two years, $\log(\# \text{answers}_{u(a),\theta,t(a)})$. Control variables include $\log (\# \text{answers}_{q(a)})$ and $\log \left(\sum_{\alpha \in A_q}\# \text{votes}_{\alpha} +1\right)$, the total number of answers provided to question $q(a)$, and the sum of all votes across these answers. To avoid problems due to $\log 0$ values, we add 1 to counts that can evaluate to 0. The plot shows point estimates with their 95\% confidence intervals. Orange markers refer regression analyses of whether or not an answer is the top answer, purple markers to analyses of the number of votes. \textbf{b.} Estimated probability  of acquiring new skills at different values of \emph{density}, users' relatedness-weighted expertise in other skills: $d_{\theta,u} = \sum_\kappa \frac{\mathrm{R}_{\theta,\kappa}}{\sum_\tau \mathrm{R}_{\theta,\tau}}   X_{\kappa,u}$, where $X_{\kappa,u}$ denotes user $u$'s prior expertise in skill $\kappa$ and $\mathrm{R}_{\theta,\kappa}$ the relatedness between skills $\theta$ and $\kappa$. \textbf{c.} Effect of skill value on the probability that a user acquires a skill from a regression analysis of whether or not a user will acquire a new skill $\theta$, on the skill's value, $V_\theta$, controlling for the user's density around the skill, $d_{\theta,u}$. Orange markers refer to analyses that use Python-related questions only, purple markers to the full sample.} \label{fig:fig_userlevel}
\end{figure}

Fig.~\ref{fig:fig_userlevel}b shows how these estimated probabilities rise with the user's expertise density around a skill. Users are most likely to pick up skills closely related to those they already master, with the probability of acquiring a new skill rising from 0.9\% (when the skill is unrelated) to 14.3\% (when it is highly related).

Apart from a close alignment with their existing skills, higher rewards may also matter. We therefore next regress a dichotomous variable that encodes skill acquisition on the skill value estimated in eq.~(\ref{eq:skill_value}) and the user's density around the skill:
\begin{equation}\label{eq:entryregression_skill_value}
y_{u,\theta,t} = \beta_V \log(V_{\theta}) + \beta_DD_{u,\theta,t} + \sigma_{u,\theta,t},
\end{equation}
where $y_{u,\theta,t}$ is a binary variable that encodes whether or not user $u$ acquires skill $\theta$ in  year $t$, $\log(V_{\theta})$ the logarithm of the skill's value, and $D_{u,\theta,t}$ the user $u$'s density around skill $\theta$ in the two years preceding year $t$, as defined in eq.~(\ref{eq:density_user}). We estimate this model once with data from all programming languages and once with data that rely only on answer posts related to Python.

%Figure~\ref{fig:fig_userlevel}a illustrates how SO activity shifted across software expertise areas between 2009 and 2022. Notable growth appears in \emph{Web Frameworks}, certain \emph{Android}-related skills, \emph{Cloud Computing}, and \emph{AI/ML}. Conversely, skills associated with \emph{Advanced Programming Concepts} (e.g., skills to \emph{Develop a data structure library for efficient data manipulation}, \emph{Develop a multimedia application in Windows}, or \emph{Develop a thread-safe application to prevent race conditions)}, as well as to \emph{Networking} or \emph{Enterprise and Web App Development}, have seen their relative importance on SO decline.

Findings are summarized in Fig.~\ref{fig:fig_userlevel}c, with detailed results described in Appendix~\ref{sec:entry_skill_value}. Somewhat surprisingly, developers do not universally pursue high-value skills. As shown in the figure's right-most panel, this result holds even when we control for how related these skills are to the user's current skill sets. This may reflect higher technological barriers or learning costs associated with high-value skills. Interestingly, however, Python users follow a different pattern: when we restrict the analysis to Python-related questions  (shown by purple markers), skill value has a positive effect on their decision to learn a new skill. This suggests that Python’s versatility allows its users to more readily acquire profitable expertise. This not only offers an explanation for Python's popularity, but also highlights its distinct role in the software ecosystem: as a general-purpose language, it enables users to branch into new, higher-value expertise areas more easily, altering and expanding their career trajectories.

\section{Discussion}

The growing complexity of the labor market has made measuring worker capabilities both more difficult and more important. Here we demonstrated that data on individual problem solving can be used to generate a useful, fine-grained taxonomy of skills used in software-related work, a key industry in the digital economy. These skills can be given coherent economic values that help predict wages in job advertisements and be used to guess which questions users provide the best answers to according to feedback from the community. This illustrates the validity and usefulness of the skill taxonomy inside and outside the context of SO. 

The fine-grained detail of the skill taxonomy allows us to assess the mapping of programming languages to skills. This reveals a nested landscape of skills, where ubiquitous skills are undemanding in terms of which programming language they require, but specialized skills require not custom-tailored languages but are rather implemented in generalist, highly versatile, languages.

A core insight from our results is that skills exist in a space of related capabilities, echoing network-based accounts of human capital \citep{anderson2017skill,alabdulkareem2018unpacking, stephany2024price}. Constructing a network based on SO postings that connects skills that are often possessed by the same users shows that real-world software development jobs are coherent, requiring  skills that are closely related to one another. Moreover, the topology of this network steers how programmers learn: over time SO users acquire new skills that are closely related to the ones they already possess. Another important predictor of skill acquisition is a skill's value. Somewhat surprisingly, in general, SO users branch into new skills that are of relatively low value. Presumably, these skills are easier to learn, which may explain their modest returns in the first place. However, this behavior is different for Python users, who preferentially acquire high-value skills. Our networked taxonomy of software skills thereby provided a micro-to-macro explanation for the rise of the Python programming language: Python enables users to learn more valuable skills. 

This interpretation is consistent with broader indicators of Python’s role in the ecosystem. Python ranks among the most widely used languages in major indices \citep{tiobe2026}, and developer surveys emphasize that it is used across many application areas \citep{jetbrains_psf_python_developers_survey_2024_purposes}. In addition, many AI-assisted data analysis tools solve tasks by executing Python code in a sandboxed environment \citep{openai_data_analysis_chatgpt}, which makes Python a natural interface between developers and expanding AI-related workflows. Empirical studies have shown that Python is one of the most accessible and ubiquitous languages \citep{juhasz2026software}. Taken together, these patterns fit an account in which Python lowers the cost of moving into high-value skill clusters.

%explain an important macro-system outcome: the rise of Python as the leading all-purpose programming language. Our explanation for Python's success is that it enables people to learn new skills that have higher economic value. Ultimately, our results highlight how deepening our understanding of the micro skill-structure of human capital offers new ways to interpret macro-level phenomena such as technology diffusion, wage dispersion, and the evolution of industry-wide practices.

%On the one hand, skills are well-defined as distinctive capability to perform specific kinds of work. This is demonstrated  by their distinct labor market values and the fact that effective learning happens by practice within skills \citep{arrow1962economic}. On the other hand, skills  also have meaningful relationships with each other, evidenced by their non-random co-occurrence within job ads and observed patterns of individual learning.

Our findings connect with the literature on skill complexity, labor market transitions, and the organization of work, complementing studies that have emphasized how granular human capital analysis can reveal hidden dynamics in career mobility, wage disparities, and skill complementarities \citep{autor2003skill,anderson2017skill,alabdulkareem2018unpacking,del2021occupational,stephany2024price}. Previous research however, tends to use standardized classifications such as O*NET to define skills and applies them to curated job descriptions, limiting the granularity and responsiveness of skill taxonomies. In contrast, by extracting skills directly from real-world problem-solving interactions on SO, our approach captures the immediate and practical realities of skill application at a global scale. The value of this approach is demonstrated in its ability to study and explain phenomena from the micro level (the coherence of job requirements and fine-grained individual learning trajectories) to the macro level (the mapping of skills to programming language and the rapid rise to dominance of Python) linking individual to collective learning dynamics.

We highlight three possible applications of our method and taxonomy. First we have demonstrated that microdata on activity can be fruitfully aggregated into skills to describe skill acquisition, development and transitions. Skills from any sector in which large-scale fine-grained information on individual behavior at work is available could be categorized in a similar manner. Second, in the software sector, employers can use our approach to identify skills related to work activities their employees actually need to perform, enabling better labor market matching outcomes and internal skill development programs. Finally, educators can better understand the labor market landscape of the software industry using our taxonomy.

There are several limitations to our study, which may be tackled in future work. First, we have focused exclusively on programming-related skills. However, software developers also need  many other skills, such as management and reporting capabilities. Although SO contains some postings about organizational aspects of software development, its emphasis is on technical subjects. To explore how programming skills are combined with other types of expertise in software jobs, future studies could use job ads to compare the co-occurrence of our programming skills with  different kinds of soft skills.

Second, although the question-answer structure makes SO ideal for identifying software work activities and their prerequisite skills, it may be less suited for descriptions of macro-level trends and individual career paths. In Appendix~\ref{section_reweighted}, we compare user numbers on SO to GitHub. This analysis suggests that in terms of capturing software developers, both datasets are closely aligned. In this Appendix, we therefore also rescale the activity in different programming languages such that they reflect their presence on GitHub. However, it is impossible to judge the representativity of SO at the level of individual programming skills. Moreover, recently, SO usage has strongly declined due to introduction of generative AI and advanced coding assistants \citep{del2024large}. In future work, we therefore plan to identify software skills directly in code posted on GitHub. 

Third, we opted to create a skill taxonomy that is time-invariant. However, an analysis of how programming skills change would provide a more vivid picture of the software industry. Such an analysis could also expand the size of our job ad dataset. Although sufficient for the validation exercise we conducted, a larger dataset would allow us to see how skills in software development and their value change over time. 

Notwithstanding these limitations, our study demonstrates how question-and-answer data can inform us about the evolution of  software development work by quantifying the abstract notion of a skill. Programming skills are rapidly diffusing beyond pure software development firms and understanding these skills and their evolution is pivotal to understanding the future of work in an increasingly digital economy. Our application contributes to such understanding by shedding light on novel aspects of this at high granularity, at scale, and in near real-time.  

\section*{Data Availability}
The dataset can be found at \url{https://zenodo.org/records/18669194?token=eyJhbGciOiJIUzUxMiJ9.eyJpZCI6ImY5N2UyNjZhLTRkNDMtNGJlOC1iZmU5LTdjMzA5ZGE2MjliNiIsImRhdGEiOnt9LCJyYW5kb20iOiIzOWY2Y2U5NjIxNjUzYjY2NzZiNTA2MzY5MjRhNzc3MCJ9.S3jZKbWuFVGDueoWDunQb2wcslOpcH4MB7jJGK8UhKypb1Kipov8ihYdu7nrAxtb_B7oMuUKj1hcDT6o_eE7fg}.

\section*{Declaration of generative AI use}
The authors acknowledge the use of the following AI tools: (a) ChatGPT-3.5 to extract job requirements and salary information from job ads. (b) ChatGPT-4.0 to provide descriptive labels for the skills in our skill taxonomy. These skill labels were used to allow for easy reference to skill examples in tables and in the main text and to link the job requirements extracted from job ads as detailed in point (a) above to our skill taxonomy. In the latter case, findings were corroborated using links that do not rely on the AI-generated skill labels. All prompts are provided in the main content.
% \section*{Code \& Data Availability}
% The dataset can be found at \url{https://zenodo.org/records/15719430?preview=1&token=eyJhbGciOiJIUzUxMiJ9.eyJpZCI6IjhhNmIyZTY5LWIwOTYtNGQyNC05NzI3LTMzZTc2MzhjYmI5NyIsImRhdGEiOnt9LCJyYW5kb20iOiIwYzNjYzA0ZGRlMDQ2YjA2ZjM2MWRhZDE0YzU3MmQ4OCJ9.GCe8uxtVI8V6MeGXDx3isrrbaZ9-Tejl6gS72sZ4D-mJMlULdiJpPeKZTQsN0HuR88lQfbQjLToucc9VhJYHYQ}. The code is available at \url{https://github.com/Roland-Feng/skill_space_NHB}.
\bibliographystyle{abbrvnat}
\bibliography{references}

\clearpage
\appendix
\counterwithin{figure}{section}
\counterwithin{table}{section}
\section{Stochastic block models}\label{sec:app_SBM}
SBMs represent generative models for networks, that coarsen networks by modeling groupings of nodes and the edge distributions between and within these groupings. These models typically postulate a stochastic data generating process of the network that produces communities and the edges among nodes within and between these communities. Observed networks are assumed to be sampled from this process.

More precisely, given an $N$-node network, $G$, and partition or community structure, ${\bf c}$ that assigns nodes to different groups, the probability of generating graph $G$ from its partition ${\bf c}$ is: 
\begin{equation}
P(G|\bf c,\theta),    
\end{equation}
where $\theta$ is a set of parameters that govern how nodes connect to one another, depending on the communities they belong to. 
Inference inverses this relation, such that it recovers the probability that a specific generative model is correct, given the observed network $G$.
Thus, the probability of observing the configuration, given the observed network $G$ with model parameters that observe prior probabilities $P(\theta,{\bf c})$ is:
\begin{equation}
P({\bf c}|G) = \frac{\sum_{\theta} P(G|\theta,{\bf c}) P(\theta, {\bf c})}{P(G)},
\end{equation}

Under the assumption that the generative model has only one set of parameters for network $G$, the equation simplifies to:
\begin{equation}
P({\bf c}|G) = \frac{P(G|\theta,{\bf c})P(\theta, {\bf c})}{P(G)}.
\end{equation}

In the bipartite implementation of this model, priors are chosen such that nodes of the same layer cannot link to one another~\cite{gerlach2018network}.
Applying this bipartite SBM to our question-tag network yields communities for both questions and tags. In the current paper, we are only interested in communities of tags (as opposed to questions) of which we discover 247, containing a total of 5,083 tags.

The procedure yields a community label for each tag, even when there is some ambiguity over which community fits the tag best. For instance, some tags are highly ubiquitous (e.g., \emph{pandas} in Python related questions) and therefore connect to nodes in various communities. This leads to noise in the community composition. To reduce this noise, we try to remove tags that are only weakly connected to the other tags in their community. In principle, SBMs can identify such nodes using MCMC simulations. However, the size of our question-tag matrix renders this approach computationally infeasible. Therefore, we here use a simpler approach.

In particular, we construct a weighted unipartite tag network, where edge weights denote the intensity with which  tags co-occur on the same questions. Next, we label nodes in this tag network according to the community structure we discovered above. For each tag, we now calculate the overrepresentation, $O_{tc}$, of (weighted) links to nodes in each of the 247 communities:
\begin{equation}
O_{tc} = \frac{w_{tc}/\sum_{c'}w_{tc'}}{\sum_{t'}w_{t'c}/\sum_{t',c'}w_{t'c'}},
\end{equation}
where $w_{tc}$ is the sum of edge weights across all links between tag $t$ and community $c$. 

$O_{tc}$ measures how important tag $t$ is to community $c$ compared to all other communities. High values of $O_{ct}$ indicate that the tag is a core part of the community, while low $O_{ct}$ values signal that the tag may be misplaced or only weakly related to community $c$. To focus on core tags, for each community, we remove the $20\%$ of tags with the lowest $O_{ct}$ values. Finally, we also remove communities with fewer than 3 tags, for which the tag composition provides too little information on the associated skill. In this way, we arrive at a skill space with 237 skills, composed of a total of 4,054 tags.

\section{Skill space}\label{sec:skill_space_labels}
In this section, we provide a detailed, annotated version of the skill space depicted in Fig.~\ref{fig:fig1_skilldef}b. This annotated skill space is shown in Fig.~\ref{fig:_space_all_label}. 

Each node is labeled with an example of a programming skill provided by ChatGPT-4.0. To do so, we prompt the LLM as follows:

\begin{quote}
    I have a list of tags related to software engineering and computer science. The tags are ordered by their importance, listing the most important tags first. Using no more than 10 words, could you please suggest a general task related to the following tags 'task\_description' as key words? Please make sure that the task is  general and not too specific. Please only return json content and not anything else.
\end{quote}

Because questions describe hands-on coding problems, this prompt refers to tasks (generalized problems), which we regard as closely related to the skills they require. Doing so avoids that the LLM tries to guess which skills are required to solve the problem stated in a question. 

\begin{figure}
\centering
\includegraphics[width=17cm]{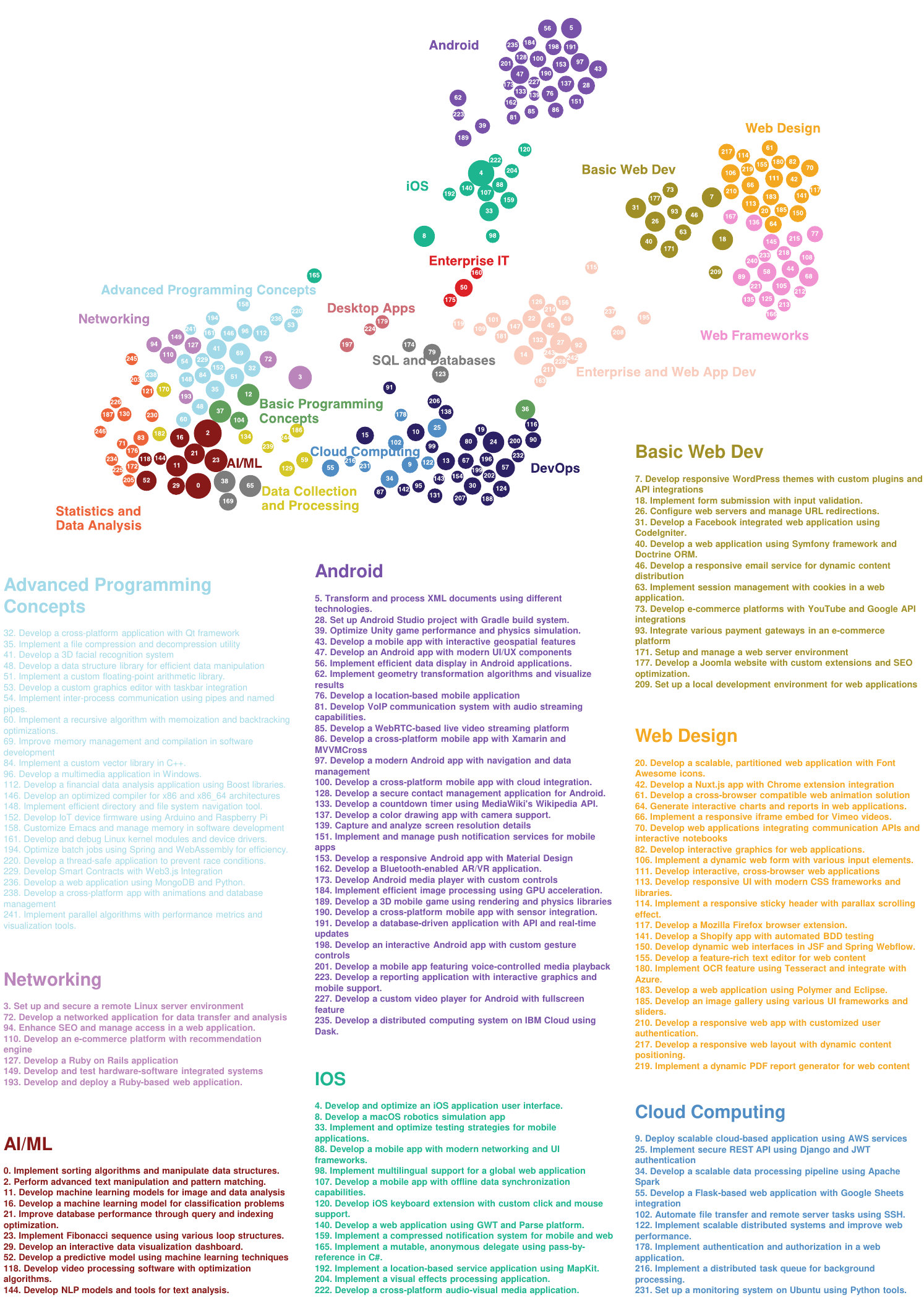}
\end{figure}
\begin{figure}
\includegraphics[width=17cm]{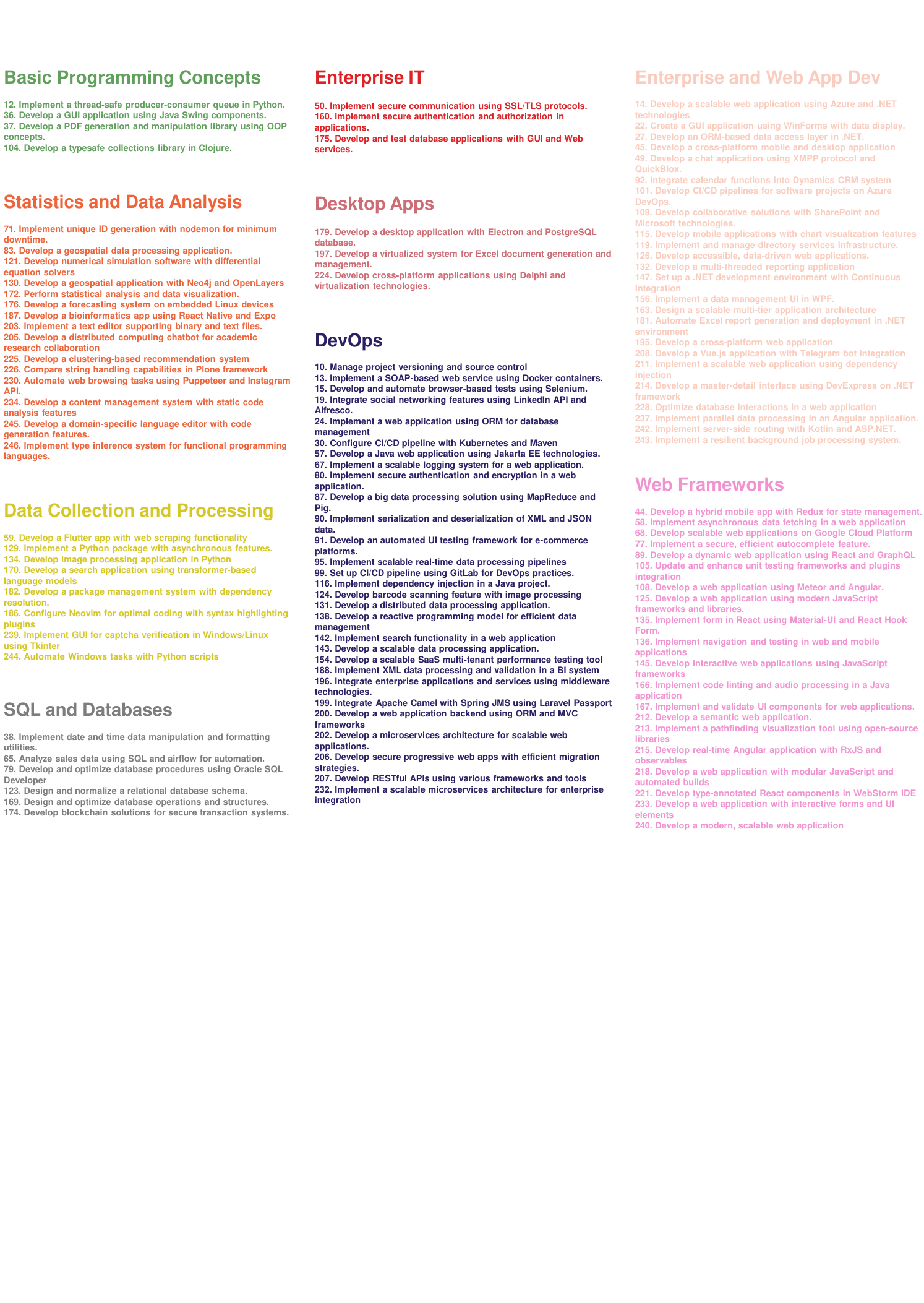}
\caption{Fully annotated skill space.}
\label{fig:_space_all_label}
\end{figure}

\begin{figure}[htbp]
    \centering
    \includegraphics[width=17cm]{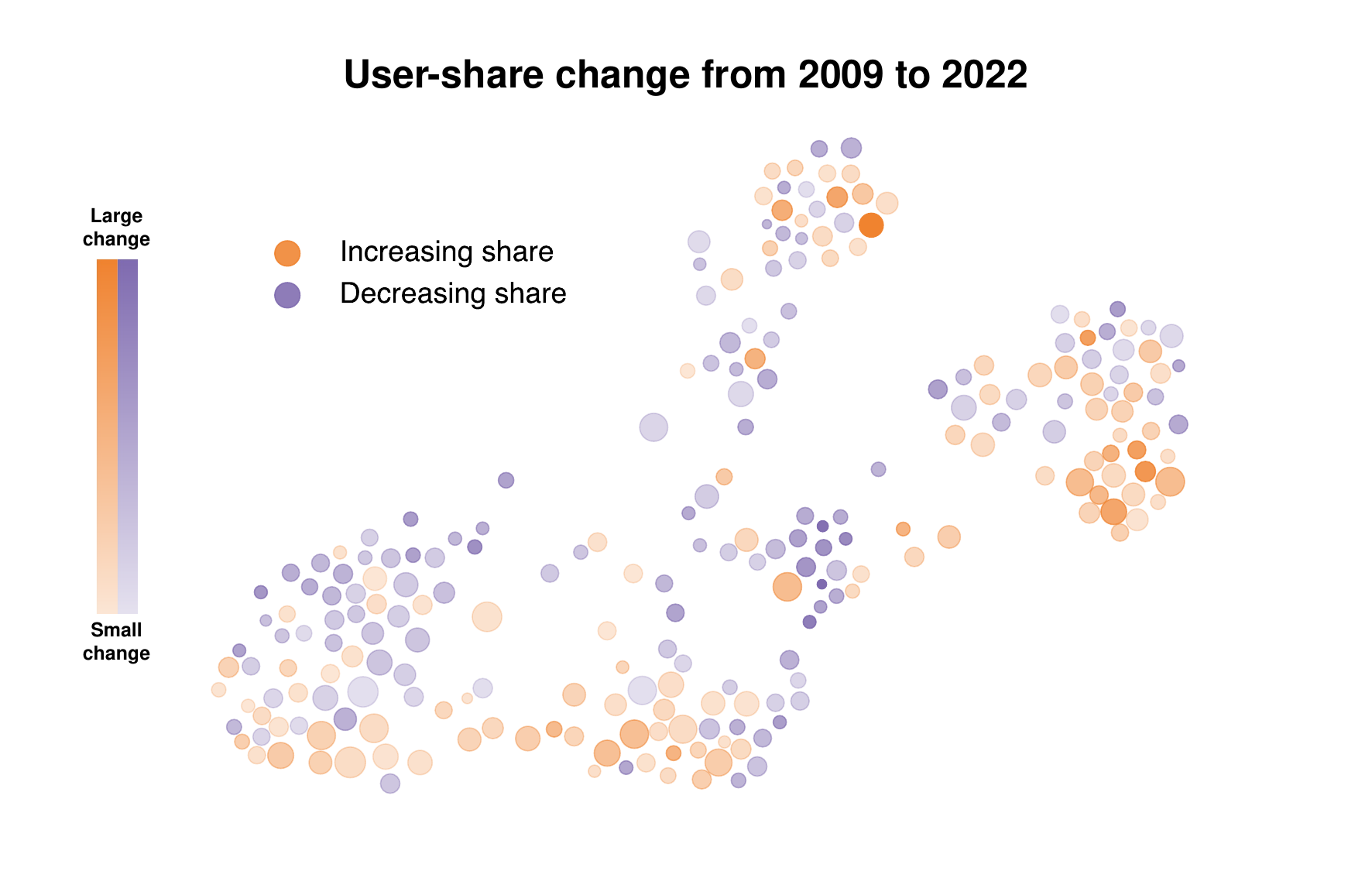}	
    \caption{
Skill dynamics. \textbf{a.} Skill user-share change from 2009 to 2022. Orange markers signal increases, purple markers decreases, in user-shares between 2009 and 2022. Marker transparency reflects the size of shifts: darker tones indicate larger changes in user shares.}\label{fig:user_share_change}
\end{figure}

Fig.~\ref{fig:user_share_change}a illustrates how SO activity shifted across software expertise areas between 2009 and 2022. Notable growth appears in \emph{Web Frameworks}, certain \emph{Android}-related skills, \emph{Cloud Computing}, and \emph{AI/ML}. Conversely, skills associated with \emph{Advanced Programming Concepts} (e.g., skills to \emph{Develop a data structure library for efficient data manipulation}, \emph{Develop a multimedia application in Windows}, or \emph{Develop a thread-safe application to prevent race conditions)}, as well as to \emph{Networking} or \emph{Enterprise and Web App Development}, have seen their relative importance on SO decline.

\section{Analyzing job advertisements}
\subsection{Example job ad}\label{sec:job_ad_example}
The analyses in Fig.~\ref{fig:fig2_jobads} of the main text, we assign skills to job advertisement. To do so, we first feed the job ad's text to a large language model (ChatGPT-3.5) to extract strings that describe all work activities (``tasks'') and salary information from the listed job description, with the following prompt:

\begin{quote}
    Here is a job ad:\{text\}. Could you extract in a json format the following information with required json key names:
    
    1. company name, with key name `company\_name'.
    
    2. bool of whether the ad is for more than one job, with key name 'multi\_jobs'. 
    
    3. With key name 'jobs', build a list containing the information of each jobs: for each job, build a key-values pair with the job name as key name and value containing the following items: bool of whether this ad contain information of yearly salary values, with key name 'salary\_index'; location, with key name 'location'; salary, with key name 'salary' and containing the following items: if the salary is a range, please give an average value with the key name 'salary\_average', the minimum value with the key name 'salary\_min', the maximum value with the key name 'salary\_max'; if the salary is not a range, please give the value with the key name 'salary\_average'; currency of salary, with key name 'currency'; bool of whether salary includes equity, with key name 'include\_equity'; a list of skills required by the job, with key name 'skills'.
    
    Please only use the content in the advertisement and do not use other outside information.
\end{quote}

Next, we use sBert\citep{reimers-2019-sentence-bert} to embed each of the listed  work activities in a 384 dimensional space. We then do the same for the skill labels we generated in Sec.~\ref{sec:skill_space_labels}. To increase the accuracy of these embeddings we prompt ChatGPT-4.0 to produce somewhat longer labels of up to 30 words. Finally for each work activity, we pick the closest skill in our taxonomy in terms of the cosine similarity between the two embeddings. We keep only those work activities for which the matched skill's cosine similarity exceeds 0.3. 

Below, we illustrate this using an example of a concrete job advertisement. The job advertisement contains the following text:

\begin{quote}
    XXX is a home security and automation company that combines industry-leading technology and truly personal service to protect our customers. We're looking for a Devops Engineer to work alongside our small engineering team (currently around 6, depending on your definitions) building robust systems and processes. We'd like someone with strong AWS chops and experience with provisioning/config management tools (we currently use Ansible and Terraform). Some other nice points to have would be experience with managing monitoring or metric aggregation systems and PCI compliance. One notable feature to me is the transparency in the organization. The whys and hows of both the past and future are well communicated across most teams. The things we work on are clearly tied to goals, and ...
\end{quote}

The job requirements the LLM extracts, together with the skills we match to them, are listed in  Table~\ref{table:job_ad_example}. The procedure leads to plausible matches. The first job requirement, AWS (Amazon Web Services, a cloud computing service), is clearly closely related to the matched skill, which contains several tags related to scheduling tasks, such as \emph{job-scheduling} or \emph{quartz-scheduler} and several amazon-specific tags such as  \emph{aws-application-load-balancer} and \emph{amazon-athena}. Also for the second job requirement, the identified match is plausible. Ansible and Terraform are two platforms for CI/CD (Continuous Integration and Continuous Delivery/Deployment) pipelines, which help   automate the building, testing, and deployment of code. In this context, Maven, which is primarily used in Java and Kubernetes, focuses on automating the deployment of containerized applications. For the third and fourth job requirements, the similarity to their matched skills falls below the threshold value of 0.3. Yet, the job requirements and matched skills are still related, albeit somewhat more weakly. In the third job requirement, metric aggregation systems are often used to monitor CPU or network loads, which support scalable real-time data processing. Similarly, the PCI (Payment Card Industry) compliance mentioned in the fourth job requirement covers security standards required when handling debit or credit card information. This typically involves the expertise in implementing secure user authentication and authorization processes related to the matched skill.

\begin{table}[htbp]
    \caption{Matching job requirements to skills. }\label{table:job_ad_example}
        \begingroup
\centering
\resizebox{\columnwidth}{!}
{\begin{tabular}{llc}
\tabularnewline \midrule \midrule
(1) job requirements & (2) matched skill & similarity (cosine)\\
\midrule
AWS & Deploy scalable cloud-based application using AWS services & 0.7987\\
provisioning/config management tools (Ansible and Terraform) & Configure CI/CD pipeline with Kubernetes and Maven & 0.4285\\
managing monitoring or metric aggregation systems & Implement scalable real-time data processing pipelines & 0.2186\\
PCI compliance & Implement secure authentication and authorization in applications&0.2487\\
\midrule \midrule
\end{tabular}}
\par\endgroup
\begin{minipage}{\linewidth}
\vspace{0.1cm}
 \begin{small}
 Column (1): job requirements as extracted by the LLM from the job ad. Column (2): the label of the matching skills. Column (3): cosine similarity between the embedding vectors of Columns~(1) and (2). 
 \end{small}
\vspace{0.1cm}
\end{minipage}
\end{table}

\subsection{Salary versus skill value}\label{section_salary_regression}
To assess how well our skill taxonomy can be used to predict salaries in job ads, we focus on the subset of job advertisements that also provide salary information.  For these jobs, we calculate the total number of skills we match to them, as well as the average value of these skills. Next, we estimate the following regression model:

\begin{equation}\label{eq:salary_regression}
\log(w_j) = \beta_v \bar{v}_j + \beta_r \bar{r}_j + \beta_n \log(n_{j}) + \tau_{y(j)} + \varepsilon_j, 
\end{equation}
where $w_j$ is the salary or midpoint of the salary range posted for job $j$, $\bar{v}_j$ is the average of the logarithm of the skill values across all skills that job $j$ requires, $\bar{r}_j$ the average of the skill density around these skills (i.e., the coherence of the skills the job requires) and $n_j$ the number of skills that job $j$ requires. $\tau_{y(j)}$ represents  year fixed effects to control for general trends in the labor market. Finally, $\varepsilon_j$ is an error term. 

Outcomes are reported in the table of Table~\ref{tab:salary_regression}. The estimated effect effect of skill value on posted salaries is around 0.9 across all models,  suggesting that a $10\%$ increase in skill value is associated with a $9\%$ higher salary.

\begin{table}[htbp]
\caption{Salary regression in online job ads.}\label{tab:salary_regression}
\centering
\begin{tabular}{lcccccc}
   \tabularnewline \midrule \midrule
   Dependent Variable: & \multicolumn{6}{c}{$\log(salary)$}\\
   Model:                        & (1)            & (2)                   & (3)            & (4)            & (5)            & (6)\\  
   \midrule
   \emph{Variables}\\
   skill value    & 0.8734$^{***}$ &                       & 0.8522$^{***}$ &                & 0.8853$^{***}$ & 0.8669$^{***}$\\   
                                 & (0.1594)       &                       & (0.1550)       &                & (0.1591)       & (0.1550)\\   
   skill coherence        &                & 0.0775$^{**}$                & 0.0273        &                &                & 0.0253\\   
                                 &                & (0.0294)              & (0.0257)       &                &                & (0.0262)\\   
   \# skills             &                &                       &                & -0.0059 & -0.0067$^{*}$ & -0.0066$^{*}$\\   
                                 &                &                       &                & (0.0034)       & (0.0032)       & (0.0033)\\   
   \midrule
   \emph{Fixed-effects}\\
   year                     & Yes            & Yes                   & Yes            & Yes            & Yes            & Yes\\  
   \midrule
   \emph{Fit statistics}\\
   Observations                  & 4,116          & 4,116                 & 4,116          & 4,116          &4,116          &4,116\\  
   R$^2$                         & 0.12595        & 0.11509               & 0.12610        & 0.11501        & 0.12745        & 0.12756\\  
   Within R$^2$                  & 0.01367        & 0.00142  & 0.01384        & 0.00133        & 0.01536        & 0.01548\\  
   \midrule \midrule

\end{tabular}
\label{tab_salary_regression}
\begin{minipage}{\linewidth}
\vspace{0.1cm}
 \begin{small}
Robust standard errors in parentheses. p-value: ***: 0.01, **: 0.05, *: 0.1. The dependent variable is the value or midpoint of the salary range posted in a job ad, \emph{skill value} is the average of the logarithm of the skill values across all skills the job requires, \emph{skill coherence} is the average of the skill density round all skills the job requires, \emph{\# skills} is the logarithm of the number of skills the job requires.
 \end{small}
\vspace{0.1cm}
\end{minipage}
\end{table}

\subsection{Alternative skill matches and thresholds}\label{section_alternative_match}
\begin{figure}[htbp]
    \centering
    \includegraphics[width=17cm]{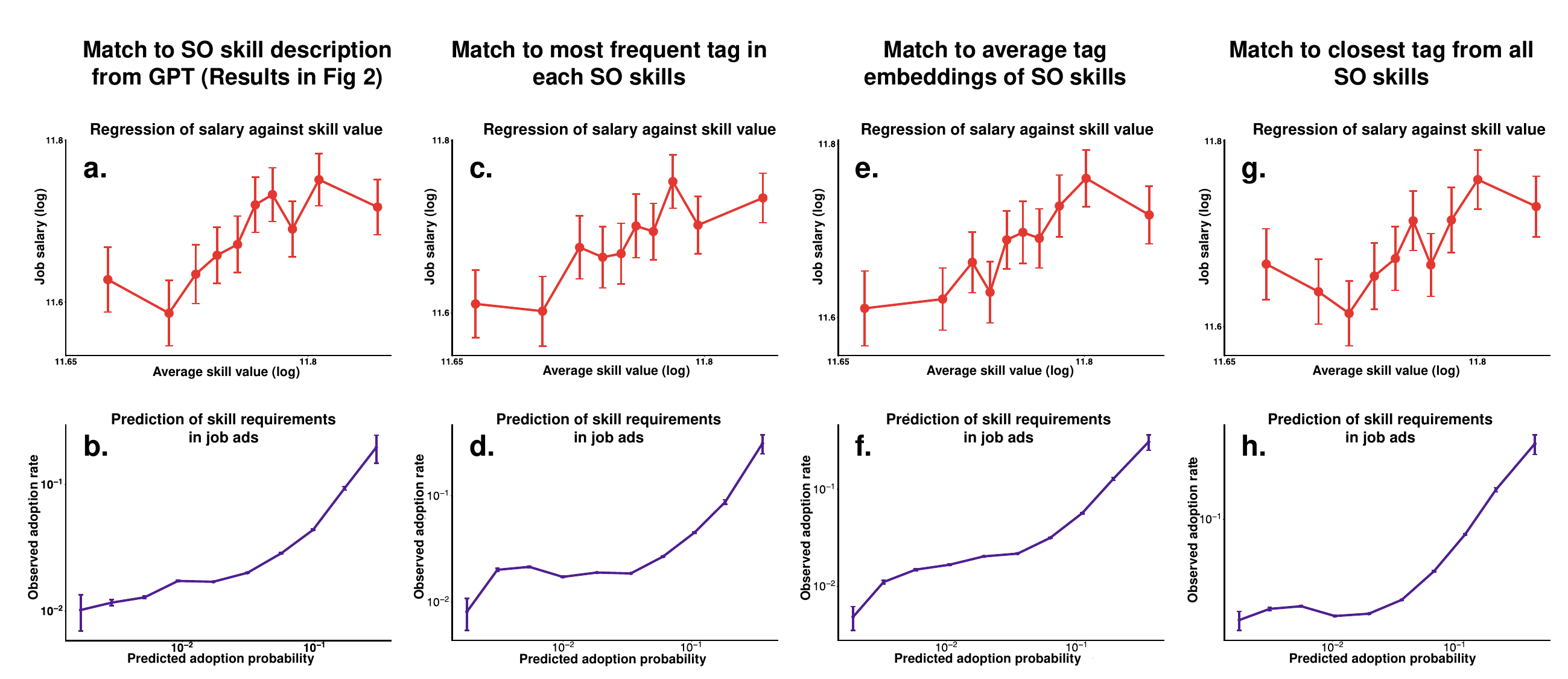}	
    \caption{Alternative methods to match job requirements in job ads to skills. \textbf{a,c,e,g.} Prediction of skill requirements. $40\%$ of skill vector elements are masked and grouped into 10 equally sized bins based on the average relatedness of the masked skill to the (unmasked) skill requirements of a job. Plot displays the estimated probability, with its $95\%$ confidence interval, that skills in a bin are required in the job. \textbf{b,d,f,h.} Predictions of wage offers. Jobs are grouped into equally sized bins based on the average value of the skills they require. Vertical axis shows the estimated mean salary posted in the job ads, with their $95\%$ confidence intervals. \textbf{a,b.} Job requirements are matched to skills using the embedding vector of the skill's label, namely the results shown in Fig~\ref{fig:fig2_jobads}. \textbf{c,d.} Job requirements are matched to skill using the embedding vector of the skill's main (most frequent) tag. \textbf{e,f.} Job requirements are matched to skill using the average embedding vector across all of the skill's tags. \textbf{g,h.} Job requirements are matched to skills using the closest embedding vector across all tags of a skill.}\label{fig:fig2_si_robust}
\end{figure}

We test three alternative methods to match job requirements in job ads to skills. To do so, let $v_t^j$ denote the embedding vector of job requirement $t$ of job $j$. Our first method, used in the main text, calculates the cosine similarity between $v_t^j$ and the embedding vector of each skill's (long) labels, and then selects the skill that yields the highest cosine similarity. Our second method uses the similarity of $v_t^j$ to the embedding of the most prominent tag (in terms of its usage on SO) in each skill. Our third method calculates the similarity of $v_t^j$ to the average embeddings of the tags defining a skill. Finally, our fourth method uses the similarity of $v_t^j$ to the embedding of the closest tag in each skill. Next, we repeat the analysis depicted in Fig.~\ref{fig:fig2_jobads}  with each of these methods. Fig.~\ref{fig:fig2_si_robust} presents the results.

\begin{figure}[htbp]
    \centering
    \includegraphics[width=12cm]{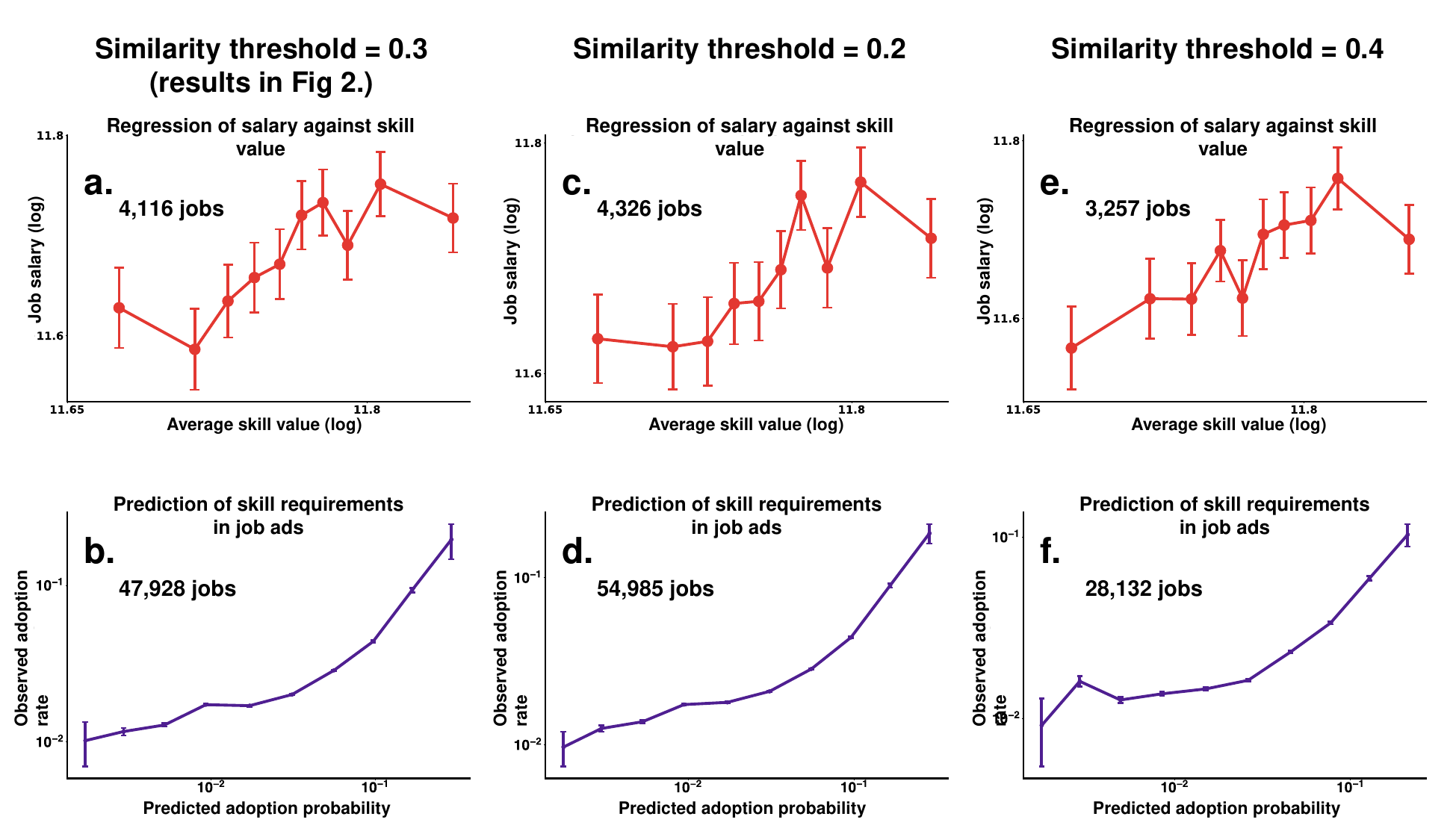}	
    \caption{Results of skill and salary predictions at varying matching thresholds.  \textbf{a,c,e.} Probability that a masked skill is required in a job. Masked skills representing $40\%$ of skill vector elements are grouped into 10 equally sized bins based on the density around the masked skill of unmasked skill requirements in a job. Plot displays estimated probabilities in each bin. \textbf{b,d,f} Predictions of salaries. Jobs are grouped into equally sized bins based on the average value of their required skills. Vertical axis shows the estimated means in each bin of the advertised wage offers. \textbf{a,b.} Job requirements are matched to skills with a minimum cosine similarity of $0.3$, namely the results shown in Fig.~\ref{fig:fig2_jobads}. \textbf{c,d.} Job requirements are matched to skills with a minimum cosine similarity of $0.2$. \textbf{e,f.} Job requirements are matched to skills with a minimum cosine similarity of $0.4$. Vertical bars represent 95\% confidence intervals.}\label{fig:fig2_si_robust_threshold}
\end{figure}

In the main text, we only keep job requirements that can be matched to skills at a cosine similarity of $0.3$ or above. To explore the significance of this threshold, we repeat our analyses with two another thresholds: $0.2$ and $0.4$. The results are shown in Fig.~\ref{fig:fig2_si_robust_threshold}. Varying the threshold around the $0.3$ value of the main text does not meaningfully alter the outcomes of this analysis: density around masked skills strongly predicts which other skills a job demands and salaries posted in job ads rise with the average values of the skill we match to them. These alternative matching methods and thresholds show that outcomes are robust and do not depend on the LLM-generated skill labels.

\section{Analysis on SO users}

\subsection{Alternative specifications for voting regression}\label{sec:alternative_voting}
In Fig.~\ref{fig:fig_userlevel}a of the main text,  we study how the popularity of an answer relates to the skill expertise of the user that provided it. The figure shows that answers by users with more skill-specific expertise tend to receive more upvotes and are more likely to be the top answer to a question. Table~\ref{tab:regression_vote_top} provides the full set of regression results related to this analysis. 

\begin{table}[htbp]
\caption{Voting regressions.}\label{tab:regression_vote_top}
        \begingroup
\centering
\begin{tabular}{lcccc}
   \tabularnewline \midrule \midrule
   Dependent Variables: & \multicolumn{2}{c}{top answer (binary)} & \multicolumn{2}{c}{\# votes (log)}\\
   Model:                       & (1)             & (2)             & (3)             & (4)\\  
   \midrule
   \emph{Variables}\\
   skill expertise      & 0.0097$^{***}$  & 0.0104$^{***}$  & 0.0130$^{***}$  & 0.0137$^{***}$\\   
                                & (0.0008)        & (0.0002)        & (0.0016)        & (0.0003)\\   
   \# answers   & -0.4886$^{***}$ & -0.4915$^{***}$ & -0.7939$^{***}$ & -0.7982$^{***}$\\   
                                & (0.0036)        & (0.0019)        & (0.0199)        & (0.0037)\\   
   \# votes all answers     & -0.0400$^{***}$ & -0.0402$^{***}$ & 0.8604$^{***}$  & 0.8596$^{***}$\\   
                                & (0.0023)        & (0.0006)        & (0.0037)        & (0.0011)\\   
   \midrule
   \emph{Fixed effects}\\
   year                         & Yes             &                 & Yes             & \\  
   skill-year            &                 & Yes             &                 & Yes\\  
   \midrule
   \emph{Fit statistics}\\
   Observations                 & 7,984,391       & 7,984,391       & 7,984,391       & 7,984,391\\  
   R$^2$                        & 0.28841         & 0.28936         & 0.81479         & 0.81518\\  
   Within R$^2$                 & 0.27190         & 0.26689         & 0.80879         & 0.80359\\  
   \midrule \midrule
   \multicolumn{5}{l}{\emph{p-value: ***: 0.01, **: 0.05, *: 0.1}}\\
\end{tabular}
\par\endgroup
\begin{minipage}{\linewidth}
\vspace{0.1cm}
 \begin{small}
Regression model of eq.~(\ref{eq:votingregression}). The dependent variable is either a binary variable that encodes whether or not an answer has received the most upvotes among all answers provided to a question, or the logarithm of the number of upvotes the answer received. \emph{skill expertise}: logarithm of skill expertise, the total number of answers the users provided to questions related to the focal skill in the past two calendar years. \emph{\# answers}: logarithm of the total number of answers provided to the focal answer's question. \emph{\# votes all answers}: logarithm of the total number of votes received by these answers. Regression models contain either year or skill-year fixed effects.
 \end{small}
\vspace{0.1cm}
\end{minipage}
\end{table}

Because for questions that receive only a single answer, this answer is automatically the top answer, we repeat the analysis in a sample that excludes any question that receives only one answer. Table~\ref{tab:regression_vote_top_2} shows that this leads to larger observed effects of skill expertise.

\begin{table}[htbp]
\caption{Voting regressions when questions receive at least two answers.}\label{tab:regression_vote_top_2}
        \begingroup
\centering
\begin{tabular}{lcccc}
   \tabularnewline \midrule \midrule
   Dependent Variables: & \multicolumn{2}{c}{top answer (binary)} & \multicolumn{2}{c}{\# votes (log)}\\
   Model:                       & (1)             & (2)             & (3)             & (4)\\  
   \midrule
   \emph{Variables}\\
   skill expertise      & 0.0260$^{***}$  & 0.0279$^{***}$  & 0.0313$^{***}$  & 0.0336$^{***}$\\   
                                & (0.0009)        & (0.0004)        & (0.0026)        & (0.0007)\\   
   \# answers    & -0.2437$^{***}$ & -0.2491$^{***}$ & -0.6309$^{***}$ & -0.6371$^{***}$\\   
                                & (0.0043)        & (0.0021)        & (0.0216)        & (0.0036)\\   
   \# votes all answers     & -0.0814$^{***}$ & -0.0816$^{***}$ & 0.7513$^{***}$  & 0.7503$^{***}$\\   
                                & (0.0065)        & (0.0011)        & (0.0046)        & (0.0014)\\   
   \midrule
   \emph{Fixed-effects}\\
   year                         & Yes             &                 & Yes             & \\  
   skill-year            &                 & Yes             &                 & Yes\\  
   \midrule
   \emph{Fit statistics}\\
   Observations                 & 3,170,868       & 3,170,868       & 3,170,868       & 3,170,868\\  
   R$^2$                        & 0.09076         & 0.09334         & 0.67850         & 0.67968\\  
   Within R$^2$                 & 0.08081         & 0.08076         & 0.66944         & 0.66110\\  
   \midrule \midrule
   \multicolumn{5}{l}{\emph{p-value: ***: 0.01, **: 0.05, *: 0.1}}\\
\end{tabular}
\par\endgroup
\begin{minipage}{\linewidth}
\vspace{0.1cm}
 \begin{small}
Idem Table~\ref{tab:regression_vote_top}, but excluding answers to questions that receive only a single answer.
 \end{small}
\end{minipage}
\end{table}

\subsection{Causal effects: Instrumental variable estimation}\label{sec:SI_causal}

In the main text, we raise the issue of sample selection bias in the regression analysis of the number of upvotes an answer receives on SO. There are two concerns. First, users can observe all existing answers to a question and, arguably, users will take these answers into consideration before deciding whether or not to provide an answer themselves. As a consequence, the answers we observe on SO are no random sample of user-answer combinations. For instance, we expect that users who believe they have little to add to existing answers to a given question will tend to refrain from posting their own answers. This bias is addressed by only focusing on the first answer (in a temporal sense) that each question received.

Second, even the first answer to a question will not be randomly sampled from a population of users and the answers they would have provided had they chosen to do so. Instead, answers will be provided by users who believe they have useful insights to the problem at hand. To address the sample selection bias this causes, we will exploit the fact that, due to timezone differences, most users are active on the SO platform at different, yet somewhat predictable, times. Therefore, we expect that who will provide the first answer to a question to some extent depends on the time of day the question was posted. This introduces some exogenous variation in the likelihood that a given user provides an answer to a given question. Moreover, as far as skill expertise is unequally distributed across timezones, this offers exogenous variation in the skill expertise associated with the first answer to a question.

We proceed as follows. First, we label each minute of a day, henceforth \emph{day-minute}, from 0 to 1399 in a manner that is independent of timezones. Next, for each user, we calculate the average minute of the day they post answers on SO. This offers a coarse indication of when a user is active on the platform. Finally, we calculate for each skill and each day-minute the amount of skill expertise that is expected to be available on SO. To do so, we sum the skill expertise of all users, weighted by the gap between the day-minute of a question and the average day-minute of each user. We call this variable, $M_{\theta, m}$, the \emph{minute-skill count} of skill $\theta$ at day-minute $m$. Next, we use this variable as an instrument for the actual skill expertise of whichever user provides the first answer to the question at hand. 

Note that there are some potential concerns about this instrument. First, if there are geographic components to skill expertise of users and if these geographic components directly influence the quality of an answer, the instrument may be invalid. For instance, the timezone that includes Silicon Valley may have a disproportionate number of highly skilled programmers, such that any questions posed when most people in this timezone are active may attract high quality first answers. To correct for this, we always include day-minute fixed effects and in some specifications also add skill--year-combination fixed effects. 

Second, the instrument assumes that questions are answered within a day from the time they are posted on SO. In fact, we would expect that our instrument does not work for questions whose first answer does not arrive within the first 24 hours after the question was raised. We exploit this in placebo tests that focus on questions whose first answer is provided at least 24 hours after the posting of the question itself.

Table~\ref{tab:regression_vote_top_IV} shows results for an adapted version of the regression model of eq.~(\ref{eq:votingregression}) in the main text, but now using Two-Stage Least Squares (2SLS) estimation. In particular, we estimate the following equation:
\begin{align}\label{eq:votingregression_IV}
y_{a,\theta} = \beta_x \log(X_{\theta,u(a)} + 1) + \beta_a \log(A_{q(a)}) + \beta_v \log(1+\sum_{i \in Q_{q(a)}} v_i) + \mu_{q(a)} + \eta_a,
\end{align}
where $u(a)$ is the user who posts answer $a$, $X_{\theta,u(a)}$ user $u(a)$'s skill-specific expertise in terms of the number of answers provided to questions involving skill $\theta$ in the two calendar years preceding the calendar year in which answer $a$ was provided, $A_{q(a)}$ the number of answers that will eventually be provided to question $q(a)$, $Q_a$ the set of these answers, $v_i$ the number of upvotes that answer $i$ received and $\eta_a$ a disturbance term. Compared to eq.~(\ref{eq:votingregression}) in the main text, we add day-minute fixed effects,  $\mu_{(q_a})$, that control for any timezone specific confounders.

As before, the dependent variable $y_{a,\theta}$ is either a binary variable that indicates whether answer $a$ is the top answer to its question or $\log(v_a +1)$, the logarithm of the number of upvotes answer $a$ receives, augmented by 1 to avoid $\log(0)$ issues. If an answer is associated with multiple skills, we replicate the observation accordingly.

The results is presented in Table~\ref{tab:regression_vote_top_IV}. The effect of skill-specific expertise on answer popularity is significant and positive  across all model specifications. Moreover, the estimated causal effects arrived at by instrumental variables estimation are substantially larger than the OLS estimates reported in the main text. This is in line with a downward sample-selection bias in the OLS results.

\begin{table}[htbp]
        \caption{Instrumental variable regression.}\label{tab:regression_vote_top_IV}
        \begingroup
        \centering
        \begin{tabular}{lcccc}
           \tabularnewline \midrule \midrule
           Dependent Variables: & \multicolumn{2}{c}{top answer (binary)} & \multicolumn{2}{c}{\# votes (log)}\\
           Model:                       & (1)             & (2)             & (3)             & (4)\\  
           \midrule
           \emph{Variables}\\
           skill expertise      & 0.0162$^{***}$  & 0.0601$^{***}$  & 0.0306$^{***}$  & 0.0919$^{***}$\\   
                                        & (0.0005)        & (0.0179)        & (0.0006)        & (0.0209)\\   
           \# answers   & -0.4847$^{***}$ & -0.4598$^{***}$ & -0.7835$^{***}$ & -0.7484$^{***}$\\   
                                        & (0.0006)        & (0.0114)        & (0.0008)        & (0.0133)\\   
           \# votes all answers    & -0.0418$^{***}$ & -0.0536$^{***}$ & 0.8555$^{***}$  & 0.8385$^{***}$\\   
                                        & (0.0002)        & (0.0048)        & (0.0003)        & (0.0056)\\   
           \midrule
           \emph{Fixed-effects}\\
           year                         & Yes             &                 & Yes             & \\  
           QA-abs-minute                 & Yes             & Yes             & Yes             & Yes\\  
           skill-year            &                 & Yes             &                 & Yes\\  
           \midrule
           \emph{Fit statistics}\\
           Observations                 & 7,984,391       & 7,984,391       & 7,984,391       & 7,984,391\\  
           R$^2$                        & 0.28734         & 0.22128         & 0.81305         & 0.78261\\  
           Within R$^2$                 & 0.27051         & 0.19626         & 0.80680         & 0.76881\\  
           \midrule \midrule
           \multicolumn{5}{l}{\emph{Clustered (group\_skill\_qminute) standard-errors in parentheses}}\\
           \multicolumn{5}{l}{\emph{Signif. Codes: ***: 0.01, **: 0.05, *: 0.1}}\\
        \end{tabular}
        \par\endgroup
\begin{minipage}{\linewidth}
\vspace{0.1cm}
 \begin{small}
Results of eq.~(\ref{eq:votingregression_IV}) estimated with 2SLS. The dependent variable is either a binary variable that encodes whether or not an answer has received the most upvotes among all answers provided to a question, or the logarithm of the number of upvotes the answer received. \emph{skill expertise}: logarithm of skill expertise, the total number of answers the user provided to questions related to the focal skill in the past two calendar years. \emph{\# answers}: logarithm of the total number of answers provided to the focal answer's question. \emph{\# votes all answers}: logarithm of the total number of votes received by these answers. Regression models contain day-minute fixed effects (\emph{QA-abs-minute}) for the timing of the question and year or -year fixed effects. To overcome sample-selection biases, we instrument skill expertise with $M_{\theta, m}$, the estimated amount of available skill expertise at the moment the question was posted, calculated as the weighted sum of skill expertise across all users, where weights reflect the temporal distance between the day-minute of a user and of the question's posting.
 \end{small}
\end{minipage}  
\end{table}

Next, we test this strategy using our placebo tests. To do so, we split the sample of questions into questions that receive a first answer within the first 24 hours after they were posted and those that receive their first answer later than that. Results are shown in Table~\ref{tab:regression_vote_top_IV_placebo}. For answers provided within 24 hours, skill expertise has a strong and statistically significant positive effect on answer popularity, regardless of whether we look at the probability of being the top answer or the number of upvotes an answer receives. In contrast, for the placebo test, comprising of answers provided more than 24 hours after the question was posted, the effect of skill expertise becomes insignificant in three out of four models. In the model without skill fixed effects, the effect on the number of votes an answer receives turns significantly negative. This suggests that the skill fixed effects control for some relevant confounders. Therefore, our preferred estimates are derived from models that include these fixed effects. These models suggest a large and positive effect of skill expertise on the popularity of answers, corroborating our analysis in the main text.

\begin{table}[htbp]
        \caption{Instrumental variable regression: placebo tests.}\label{tab:regression_vote_top_IV_placebo}
\begingroup
\centering
\resizebox{\columnwidth}{!}{
\begin{tabular}{lcccccccc}
   \tabularnewline \midrule \midrule
   Dependent Variables: & \multicolumn{2}{c}{top answer (binary, $\leq\text{24 hours}$)} & \multicolumn{2}{c}{\# votes (log, $\leq\text{24 hours}$)} & \multicolumn{2}{c}{top answer (binary, $\geq\text{24 hours}$)} & \multicolumn{2}{c}{\# votes (log, $\geq\text{24 hours}$)}\\
   Model:                       & (1)             & (2)             & (3)             & (4)             & (5)             & (6)             & (7)             & (8)\\  
   \midrule
   \emph{Variables}\\
   skill expertise      & 0.0183$^{***}$  & 0.0620$^{***}$  & 0.0314$^{***}$  & 0.0947$^{***}$  & 0.0009          & 0.0561          & -0.0172$^{***}$ & -0.0030\\   
                                & (0.0005)        & (0.0174)        & (0.0006)        & (0.0203)        & (0.0038)        & (0.2365)        & (0.0048)        & (0.2827)\\   
   \# answers    & -0.4800$^{***}$ & -0.4528$^{***}$ & -0.7750$^{***}$ & -0.7347$^{***}$ & -0.5116$^{***}$ & -0.4920$^{***}$ & -0.8835$^{***}$ & -0.8773$^{***}$\\   
                                & (0.0007)        & (0.0123)        & (0.0008)        & (0.0143)        & (0.0023)        & (0.0848)        & (0.0032)        & (0.1014)\\   
   \# votes all answers     & -0.0448$^{***}$ & -0.0573$^{***}$ & 0.8471$^{***}$  & 0.8284$^{***}$  & -0.0231$^{***}$ & -0.0275         & 0.9150$^{***}$  & 0.9140$^{***}$\\   
                                & (0.0002)        & (0.0050)        & (0.0004)        & (0.0058)        & (0.0004)        & (0.0190)        & (0.0007)        & (0.0227)\\   
   \midrule
   \emph{Fixed-effects}\\
   year                         & Yes             &                 & Yes             &                 & Yes             &                 & Yes             & \\  
   QA-abs-minute                 & Yes             & Yes             & Yes             & Yes             & Yes             & Yes             & Yes             & Yes\\  
   skill-year            &                 & Yes             &                 & Yes             &                 & Yes             &                 & Yes\\  
   \midrule
   \emph{Fit statistics}\\
   Observations                 & 6,915,424       & 6,915,424       & 6,915,424       & 6,915,424       & 1,068,555       & 1,068,555       & 1,068,555       & 1,068,555\\  
   R$^2$                        & 0.28071         & 0.21444         & 0.80404         & 0.77090         & 0.32491         & 0.24001         & 0.86829         & 0.87025\\  
   Within R$^2$                 & 0.26317         & 0.18891         & 0.79722         & 0.75538         & 0.31531         & 0.22360         & 0.86502         & 0.86413\\  
   \midrule \midrule
   \multicolumn{9}{l}{\emph{Clustered (group\_skill\_qminute) standard-errors in parentheses}}\\
   \multicolumn{9}{l}{\emph{p-value: ***: 0.01, **: 0.05, *: 0.1}}\\
\end{tabular}}
\par\endgroup
\begin{minipage}{\linewidth}
\vspace{0.1cm}
 \begin{small}
Idem Table~\ref{tab:regression_vote_top_IV}, but splitting the sample into questions that were answered within 24 hours and those that weren't. The results for the latter sample, depicted in the in columns~(5)-(8), represent placebo tests for the instrumental variables estimation.
 \end{small}
\end{minipage}  
\end{table}

\subsection{Regression of skill acquisition on skill value}\label{sec:entry_skill_value}
To study how skill values effect which new skills users learn, we run the regression described in eq.~(\ref{eq:entryregression_skill_value}) of the main text. We compare the association of two variables, skill value and skill density, with the likelihood that users acquire a certain new skoill. We repeat this analysis once using all answers to construct expertise and skill acquisition variables, and once using only answers to Python related questions.  The results are shown in Table~\ref{tab:regression__value}.

\begin{table}[htbp]
        \caption{Skill acquisition regressions}\label{tab:regression__value}
\begingroup
\centering
\resizebox{\columnwidth}{!}{
\begin{tabular}{lcccccccc}
   \tabularnewline \midrule \midrule
   Dependent Variable: & \multicolumn{4}{c}{skill acquisition (binary, all programming languages)}& \multicolumn{4}{c}{skill acquisition (binary, Python)}\\
   Model:                           & (1)                  & (2)                   & (3)             & (4)             & (5)            & (6)            & (7)            & (8)\\  
   \midrule
   \emph{Variables}\\
   skill value           & -0.0078$^{***}$      & -0.0079$^{***}$       & -0.0068$^{***}$ & -0.0069$^{***}$ & 0.0342$^{***}$ & 0.0348$^{***}$ & 0.0140$^{***}$ & 0.0146$^{***}$\\   
                                    & (0.0008)             & (0.0008)              & (0.0006)        & (0.0006)        & (0.0011)       & (0.0011)       & (0.0009)       & (0.0011)\\   
   density    &                      &                       & 0.0194$^{***}$  & 0.0173$^{***}$  &                &                & 0.0181$^{***}$ & 0.0176$^{***}$\\   
                                    &                      &                       & (0.0002)        & (0.0002)        &                &                & (0.0004)       & (0.0006)\\   
   \midrule
   \emph{Fixed-effects}\\
   user                         &                      & Yes                   &                 & Yes             &                & Yes            &                & Yes\\  
   \midrule
   \emph{Fit statistics}\\
   Observations                     & 24,367,254           & 24,367,254            & 24,367,254      & 24,367,254      & 2,229,882      & 2,229,882      & 2,229,882      & 2,229,882\\  
   R$^2$                            & $1.8\times 10^{-5}$  & 0.01688               & 0.01859         & 0.02743         & 0.00047        & 0.01627        & 0.02210        & 0.03082\\  
   Within R$^2$                     &                      & $1.86\times 10^{-5}$  &                 & 0.01076         &                & 0.00049        &                & 0.01528\\  
   \midrule \midrule
\end{tabular}}
\par\endgroup
\begin{minipage}{\linewidth}
\vspace{0.1cm}
 \begin{small}
Detailed outcomes of the regression model described in eq.~(\ref{eq:entryregression_skill_value}) of the main text. Observations are user-skill combinations for which the user did not answer any questions related to the skill in the previous two years. The dependent variable is a binary variable that encodes whether or not the user acquires the skill (i.e., answers questions related to this skill) in the current period. \emph{skill value}: logarithm of the skill's imputed value, \emph{skill density}: the user's prior expertise in other skills, weighted by the relatedness of these skills to the current skill. For ease of interpretation, we subtract the variable's mean and then divide by its standard deviation. Models in columns (2), (4), (6) and (8) control for user fixed effect. Columns (1)-(4) report analyses performed using all answer posts, columns (5)-(8) use only Python related answer posts. Clustered (user) standard-errors in parentheses. p-values: ***: 0.01, **: 0.05, *: 0.1. 
 \end{small}
\end{minipage}
\end{table}

\section{Rescaling by Github}\label{section_reweighted}
\begin{figure}[htbp]
    \centering
    \includegraphics[width=17cm]{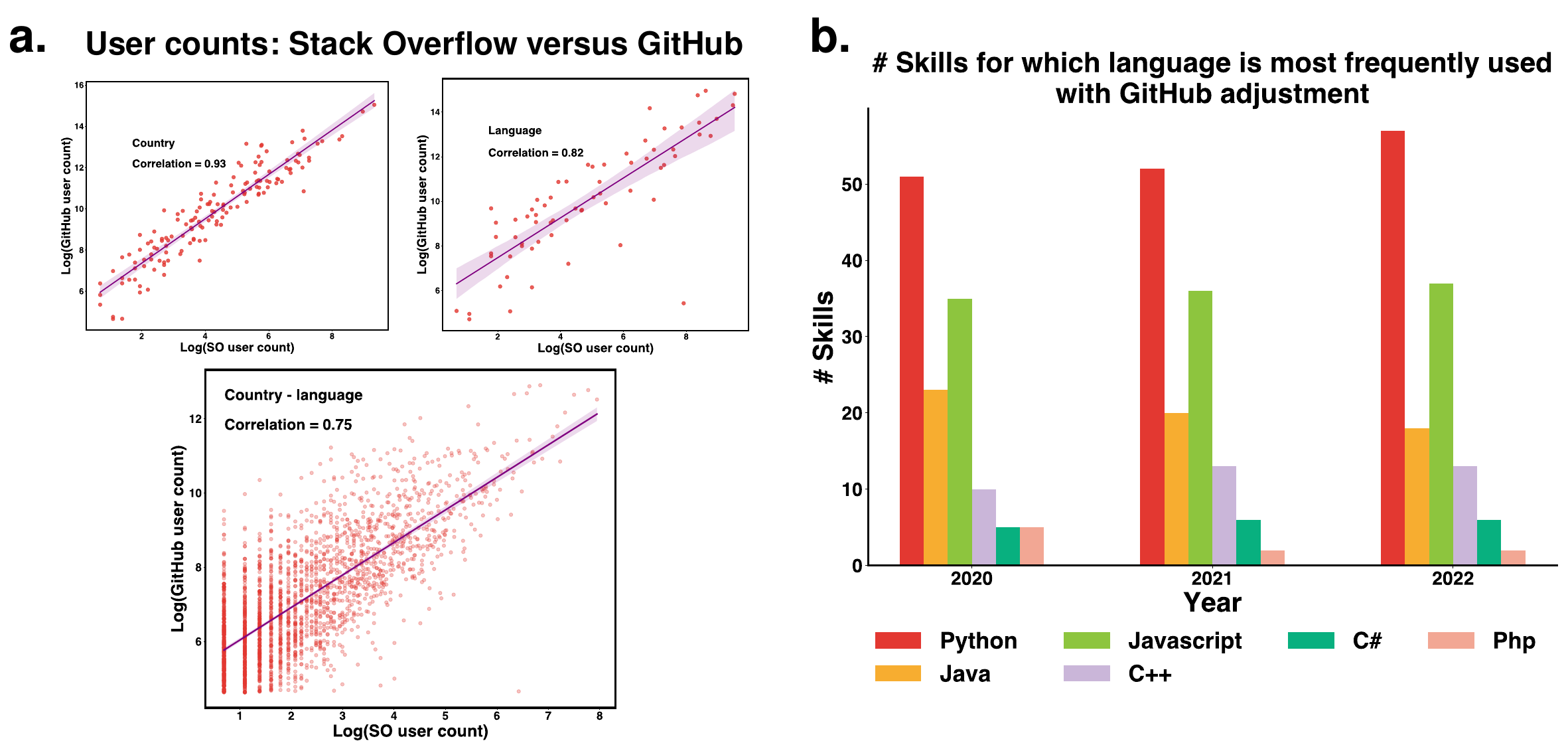}	
    \caption{\textbf{a.} The GitHub user counts against the SO user counts in the perspectives of country, programming languages and country-language in the year 2022. The values are in log scale and the $R^2$ values of the regression fit are provided inside each panel. \textbf{b.} Number of s in which a programming language is the top language in terms of SO users reweighted by user counts from GitHub, from 2020 to 2022. The graph shows the largest 6 languages measured by cumulative SO posts between August 2008 and June 2023.}\label{fig:language_change_fixed}
\end{figure}

Fig.~\ref{fig:fig_python} shows Python's progressive rise as the dominant language used in an increasing number of programming skills. However, even though SO may provide a useful dataset to define such programming skills, it is not guaranteed to be representative of programmers across the world. To assess this problem, we compare counts of programmers on SO to comparable ones on GitHub. GitHub is the world's largest software development collaboration platform and therefore offers a good representation of the global software development sector. In particular, we collect information on the (self-reported) countries of residence on GitHub and (observed) language use between 2020 and 2022~\citep{GitHub2025InnovationGraph}. Fig.~\ref{fig:language_change_fixed}a compares SO to GitHub user counts at the level of countries, languages and country-language combinations. This shows that user numbers are highly correlated across the two platforms, with correlations ranging from 0.75 to 0.93.

Nevertheless, some languages are better represented on SO than others. This may affect our analysis of how dominant a language is in a given skill. In Fig.~\ref{fig:language_change_fixed}b we attempt to correct for this, by reweighting the user counts by language using their observed shares on GitHub. This reweighting ensures that the distribution of users across language on SO mimics the one on GitHub. Because we only have GitHub information for the years 2020, 2021 and 2022, the analysis is limited to those years. Also with reweighted user counts, Python leads in most skills, albeit with around 50, in somewhat fewer skills than what we report in the main text. The ordering of the remaining languages also remains mostly unchanged, except for C\#, which was somewhat more prominent before rescaling user counts. Moreover, also the most prominent temporal patterns over the 3-year time period are mostly left unchanged. Fig.~\ref{fig:language_change_fixed}b retains the rise of Python and the drops of Java and PHP, as well as the relative stability in these years of JavaScript, validating the results shown in the main text.

\section{Swift vs Objective-C}\label{section_swift_objc}

\begin{figure}[htbp]
    \centering
    \includegraphics[width=12cm]{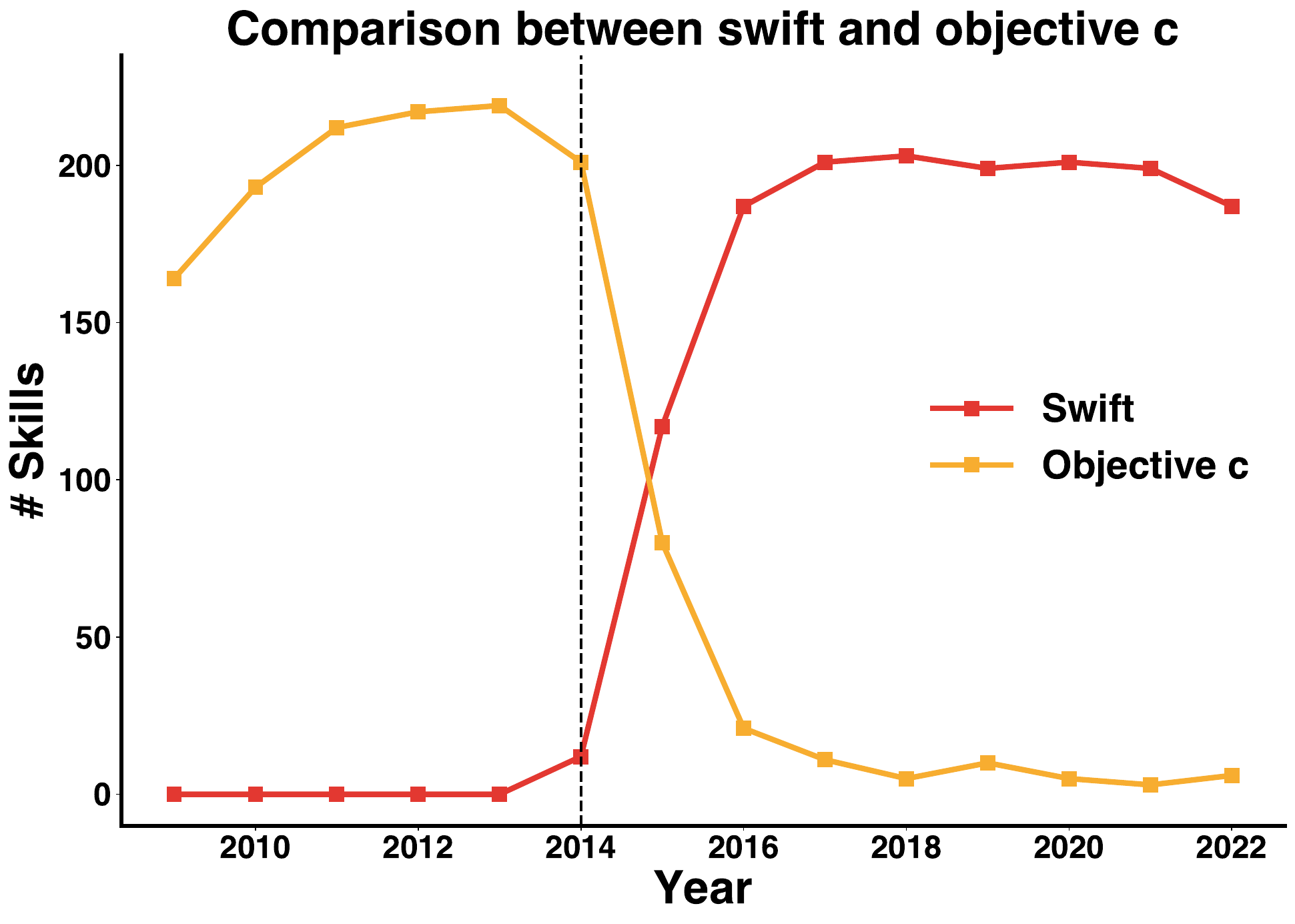}	
    \caption{Number of skills in which Objective-C has more users on SO than Swift and vice versa. The dashed vertical line marks the year 2014, the year that Swift was released.}\label{fig:fig2_si_swift_vs_objc}
\end{figure}
Software written for Apple devices was mostly written in \emph{Objective-C} from the 1980s until the mid 2010s. Apple designed \emph{Swift} to replace \emph{Objective-C} and launched the language in 2014. Within the next few years, most users switched to this new language to be able to continue developing software for Apple products ~\citep{jetbrain}. Unlike typical programming language  dynamics, which evolve organically from developer communities, Swift’s introduction was a deliberate, top-down intervention by Apple. 

Fig.~\ref{fig:fig2_si_swift_vs_objc} describes this shift in terms of the skills where Swift was more popular than Objective-C and vice versa. Before 2015, most skills were deployed preferentially using Objective-C. However, since the release of Swift in 2014 this rapidly changes. Swift surpasses Objective-C in terms of the number of different skills in which it dominates around 2015 and all but eclipses Objective-C by 2016.

\end{document}